\documentclass[11pt]{article}

\usepackage[utf8]{inputenc}           
\usepackage[english]{babel}
\usepackage{fullpage}
\usepackage{arydshln}

\usepackage{amsfonts}                 
\usepackage{amsmath}                  
\usepackage{xcolor}                   
\usepackage{tabularx}                 
\usepackage{makecell}                 
\usepackage{colortbl}                 
\usepackage{verbatim}                 
\usepackage{fancyvrb}                 
\usepackage{graphicx}                 
\usepackage{xcolor}                   
\usepackage{ragged2e}                 
\usepackage{framed}                   
\usepackage{tcolorbox}                
\usepackage{pdflscape}
\usepackage[misc]{ifsym}              
\usepackage{xspace}                   
\xspaceaddexceptions{]\}}

\usepackage[bookmarks=false]          
            {hyperref}                
\hypersetup{                          
    colorlinks,                       
    linkcolor={red!80!black},         
    citecolor={blue!65!black},        
    urlcolor={blue!80!black}          
}

\usepackage{csquotes}                 
\usepackage[backend=bibtex,maxbibnames=99,defernumbers=true]{biblatex}          
\bibliography{bibliography.bib}

\definecolor{MyColor0}{rgb}{0.940015,0.975158,0.131326}
\definecolor{MyColor1}{rgb}{0.941896,0.96859,0.140956}
\definecolor{MyColor2}{rgb}{0.944152,0.961916,0.146861}
\definecolor{MyColor3}{rgb}{0.946602,0.95519,0.150328}
\definecolor{MyColor4}{rgb}{0.949151,0.948435,0.152178}
\definecolor{MyColor5}{rgb}{0.951726,0.941671,0.152925}
\definecolor{MyColor6}{rgb}{0.954287,0.934908,0.152921}
\definecolor{MyColor7}{rgb}{0.956808,0.928152,0.152409}
\definecolor{MyColor8}{rgb}{0.959276,0.921407,0.151566}
\definecolor{MyColor9}{rgb}{0.961681,0.914672,0.15052}
\definecolor{MyColor10}{rgb}{0.964021,0.90795,0.14937}
\definecolor{MyColor11}{rgb}{0.966271,0.901249,0.14818}
\definecolor{MyColor12}{rgb}{0.968443,0.894564,0.147014}
\definecolor{MyColor13}{rgb}{0.970533,0.887896,0.145919}
\definecolor{MyColor14}{rgb}{0.97253,0.88125,0.144923}
\definecolor{MyColor15}{rgb}{0.974443,0.874622,0.144061}
\definecolor{MyColor16}{rgb}{0.976265,0.868016,0.143351}
\definecolor{MyColor17}{rgb}{0.977995,0.861432,0.142808}
\definecolor{MyColor18}{rgb}{0.979644,0.854866,0.142453}
\definecolor{MyColor19}{rgb}{0.98119,0.848329,0.142279}
\definecolor{MyColor20}{rgb}{0.982653,0.841812,0.142303}
\definecolor{MyColor21}{rgb}{0.984031,0.835315,0.142528}
\definecolor{MyColor22}{rgb}{0.985314,0.828846,0.142945}
\definecolor{MyColor23}{rgb}{0.986509,0.822401,0.143557}
\definecolor{MyColor24}{rgb}{0.987621,0.815978,0.144363}
\definecolor{MyColor25}{rgb}{0.988648,0.809579,0.145357}
\definecolor{MyColor26}{rgb}{0.989587,0.803205,0.146529}
\definecolor{MyColor27}{rgb}{0.990439,0.796859,0.14787}
\definecolor{MyColor28}{rgb}{0.991209,0.790537,0.149377}
\definecolor{MyColor29}{rgb}{0.991897,0.784239,0.151042}
\definecolor{MyColor30}{rgb}{0.992505,0.777967,0.152855}
\definecolor{MyColor31}{rgb}{0.993033,0.77172,0.154808}
\definecolor{MyColor32}{rgb}{0.993482,0.765499,0.156891}
\definecolor{MyColor33}{rgb}{0.993851,0.759304,0.159092}
\definecolor{MyColor34}{rgb}{0.994141,0.753137,0.161404}
\definecolor{MyColor35}{rgb}{0.994355,0.746995,0.163821}
\definecolor{MyColor36}{rgb}{0.994495,0.74088,0.166335}
\definecolor{MyColor37}{rgb}{0.994561,0.734791,0.168938}
\definecolor{MyColor38}{rgb}{0.994553,0.728728,0.171622}
\definecolor{MyColor39}{rgb}{0.994474,0.722691,0.174381}
\definecolor{MyColor40}{rgb}{0.994324,0.716681,0.177208}
\definecolor{MyColor41}{rgb}{0.994103,0.710698,0.180097}
\definecolor{MyColor42}{rgb}{0.993814,0.704741,0.183043}
\definecolor{MyColor43}{rgb}{0.993456,0.69881,0.186041}
\definecolor{MyColor44}{rgb}{0.993032,0.692907,0.189084}
\definecolor{MyColor45}{rgb}{0.992541,0.68703,0.19217}
\definecolor{MyColor46}{rgb}{0.991985,0.681179,0.195295}
\definecolor{MyColor47}{rgb}{0.991365,0.675355,0.198453}
\definecolor{MyColor48}{rgb}{0.990681,0.669558,0.201642}
\definecolor{MyColor49}{rgb}{0.989935,0.663787,0.204859}
\definecolor{MyColor50}{rgb}{0.989128,0.658043,0.2081}
\definecolor{MyColor51}{rgb}{0.98826,0.652325,0.211364}
\definecolor{MyColor52}{rgb}{0.987332,0.646633,0.214648}
\definecolor{MyColor53}{rgb}{0.986345,0.640969,0.217948}
\definecolor{MyColor54}{rgb}{0.985301,0.63533,0.221265}
\definecolor{MyColor55}{rgb}{0.984199,0.629718,0.224595}
\definecolor{MyColor56}{rgb}{0.983041,0.624131,0.227937}
\definecolor{MyColor57}{rgb}{0.981826,0.618572,0.231287}
\definecolor{MyColor58}{rgb}{0.980556,0.613039,0.234646}
\definecolor{MyColor59}{rgb}{0.979233,0.607532,0.238013}
\definecolor{MyColor60}{rgb}{0.977856,0.602051,0.241387}
\definecolor{MyColor61}{rgb}{0.976428,0.596595,0.244767}
\definecolor{MyColor62}{rgb}{0.974947,0.591165,0.248151}
\definecolor{MyColor63}{rgb}{0.973416,0.585761,0.25154}
\definecolor{MyColor64}{rgb}{0.971835,0.580382,0.254931}
\definecolor{MyColor65}{rgb}{0.970205,0.575028,0.258325}
\definecolor{MyColor66}{rgb}{0.968526,0.5697,0.261721}
\definecolor{MyColor67}{rgb}{0.966798,0.564396,0.265118}
\definecolor{MyColor68}{rgb}{0.965024,0.559118,0.268513}
\definecolor{MyColor69}{rgb}{0.963203,0.553865,0.271909}
\definecolor{MyColor70}{rgb}{0.961336,0.548636,0.275305}
\definecolor{MyColor71}{rgb}{0.959424,0.543431,0.278701}
\definecolor{MyColor72}{rgb}{0.957469,0.53825,0.282096}
\definecolor{MyColor73}{rgb}{0.95547,0.533093,0.28549}
\definecolor{MyColor74}{rgb}{0.953428,0.52796,0.288883}
\definecolor{MyColor75}{rgb}{0.951344,0.52285,0.292275}
\definecolor{MyColor76}{rgb}{0.949217,0.517763,0.295662}
\definecolor{MyColor77}{rgb}{0.947051,0.512699,0.299049}
\definecolor{MyColor78}{rgb}{0.944844,0.507658,0.302433}
\definecolor{MyColor79}{rgb}{0.942598,0.502639,0.305816}
\definecolor{MyColor80}{rgb}{0.940313,0.497642,0.309197}
\definecolor{MyColor81}{rgb}{0.93799,0.492667,0.312575}
\definecolor{MyColor82}{rgb}{0.93563,0.487712,0.315952}
\definecolor{MyColor83}{rgb}{0.933232,0.48278,0.319325}
\definecolor{MyColor84}{rgb}{0.930798,0.477867,0.322697}
\definecolor{MyColor85}{rgb}{0.928329,0.472975,0.326067}
\definecolor{MyColor86}{rgb}{0.925825,0.468103,0.329435}
\definecolor{MyColor87}{rgb}{0.923287,0.463251,0.332801}
\definecolor{MyColor88}{rgb}{0.920714,0.458417,0.336166}
\definecolor{MyColor89}{rgb}{0.918109,0.453603,0.339529}
\definecolor{MyColor90}{rgb}{0.915471,0.448807,0.34289}
\definecolor{MyColor91}{rgb}{0.9128,0.444029,0.346251}
\definecolor{MyColor92}{rgb}{0.910098,0.439268,0.34961}
\definecolor{MyColor93}{rgb}{0.907365,0.434524,0.35297}
\definecolor{MyColor94}{rgb}{0.904601,0.429797,0.356329}
\definecolor{MyColor95}{rgb}{0.901807,0.425087,0.359688}
\definecolor{MyColor96}{rgb}{0.898984,0.420392,0.363047}
\definecolor{MyColor97}{rgb}{0.896131,0.415712,0.366407}
\definecolor{MyColor98}{rgb}{0.89325,0.411048,0.369768}
\definecolor{MyColor99}{rgb}{0.89034,0.406398,0.37313}
\definecolor{MyColor100}{rgb}{0.887402,0.401762,0.376494}
\definecolor{MyColor101}{rgb}{0.884436,0.397139,0.37986}
\definecolor{MyColor102}{rgb}{0.881443,0.392529,0.383229}
\definecolor{MyColor103}{rgb}{0.878423,0.387932,0.3866}
\definecolor{MyColor104}{rgb}{0.875376,0.383347,0.389976}
\definecolor{MyColor105}{rgb}{0.872303,0.378774,0.393355}
\definecolor{MyColor106}{rgb}{0.869203,0.374212,0.396738}
\definecolor{MyColor107}{rgb}{0.866078,0.36966,0.400126}
\definecolor{MyColor108}{rgb}{0.862927,0.365119,0.403519}
\definecolor{MyColor109}{rgb}{0.85975,0.360588,0.406917}
\definecolor{MyColor110}{rgb}{0.856547,0.356066,0.410322}
\definecolor{MyColor111}{rgb}{0.853319,0.351553,0.413734}
\definecolor{MyColor112}{rgb}{0.850066,0.347048,0.417153}
\definecolor{MyColor113}{rgb}{0.846788,0.342551,0.420579}
\definecolor{MyColor114}{rgb}{0.843484,0.338062,0.424013}
\definecolor{MyColor115}{rgb}{0.840155,0.33358,0.427455}
\definecolor{MyColor116}{rgb}{0.836801,0.329105,0.430905}
\definecolor{MyColor117}{rgb}{0.833422,0.324635,0.434366}
\definecolor{MyColor118}{rgb}{0.830018,0.320172,0.437836}
\definecolor{MyColor119}{rgb}{0.826588,0.315714,0.441316}
\definecolor{MyColor120}{rgb}{0.823132,0.311261,0.444806}
\definecolor{MyColor121}{rgb}{0.819651,0.306812,0.448306}
\definecolor{MyColor122}{rgb}{0.816144,0.302368,0.451816}
\definecolor{MyColor123}{rgb}{0.812612,0.297928,0.455338}
\definecolor{MyColor124}{rgb}{0.809052,0.293491,0.45887}
\definecolor{MyColor125}{rgb}{0.805467,0.289057,0.462415}
\definecolor{MyColor126}{rgb}{0.801855,0.284626,0.465971}
\definecolor{MyColor127}{rgb}{0.798216,0.280197,0.469538}
\definecolor{MyColor128}{rgb}{0.794549,0.27577,0.473117}
\definecolor{MyColor129}{rgb}{0.790855,0.271345,0.476706}
\definecolor{MyColor130}{rgb}{0.787133,0.266922,0.480307}
\definecolor{MyColor131}{rgb}{0.783383,0.2625,0.483918}
\definecolor{MyColor132}{rgb}{0.779604,0.258078,0.487539}
\definecolor{MyColor133}{rgb}{0.775796,0.253658,0.491171}
\definecolor{MyColor134}{rgb}{0.771958,0.249237,0.494813}
\definecolor{MyColor135}{rgb}{0.76809,0.244817,0.498465}
\definecolor{MyColor136}{rgb}{0.764193,0.240396,0.502126}
\definecolor{MyColor137}{rgb}{0.760264,0.235976,0.505794}
\definecolor{MyColor138}{rgb}{0.756304,0.231555,0.509468}
\definecolor{MyColor139}{rgb}{0.752312,0.227133,0.513149}
\definecolor{MyColor140}{rgb}{0.748289,0.222711,0.516834}
\definecolor{MyColor141}{rgb}{0.744232,0.218288,0.520524}
\definecolor{MyColor142}{rgb}{0.740143,0.213864,0.524216}
\definecolor{MyColor143}{rgb}{0.736019,0.209439,0.527908}
\definecolor{MyColor144}{rgb}{0.731862,0.205013,0.531601}
\definecolor{MyColor145}{rgb}{0.72767,0.200586,0.535293}
\definecolor{MyColor146}{rgb}{0.723444,0.196158,0.538981}
\definecolor{MyColor147}{rgb}{0.719181,0.191729,0.542663}
\definecolor{MyColor148}{rgb}{0.714883,0.187299,0.546338}
\definecolor{MyColor149}{rgb}{0.710549,0.182868,0.550004}
\definecolor{MyColor150}{rgb}{0.706178,0.178437,0.553657}
\definecolor{MyColor151}{rgb}{0.701769,0.174005,0.557296}
\definecolor{MyColor152}{rgb}{0.697324,0.169573,0.560919}
\definecolor{MyColor153}{rgb}{0.69284,0.165141,0.564522}
\definecolor{MyColor154}{rgb}{0.688318,0.160709,0.568103}
\definecolor{MyColor155}{rgb}{0.683758,0.156278,0.57166}
\definecolor{MyColor156}{rgb}{0.67916,0.151848,0.575189}
\definecolor{MyColor157}{rgb}{0.674522,0.147419,0.578688}
\definecolor{MyColor158}{rgb}{0.669845,0.142992,0.582154}
\definecolor{MyColor159}{rgb}{0.665129,0.138566,0.585582}
\definecolor{MyColor160}{rgb}{0.660374,0.134144,0.588971}
\definecolor{MyColor161}{rgb}{0.65558,0.129725,0.592317}
\definecolor{MyColor162}{rgb}{0.650746,0.125309,0.595617}
\definecolor{MyColor163}{rgb}{0.645872,0.120898,0.598867}
\definecolor{MyColor164}{rgb}{0.640959,0.116492,0.602065}
\definecolor{MyColor165}{rgb}{0.636008,0.112092,0.605205}
\definecolor{MyColor166}{rgb}{0.631017,0.107699,0.608287}
\definecolor{MyColor167}{rgb}{0.625987,0.103312,0.611305}
\definecolor{MyColor168}{rgb}{0.620919,0.098934,0.614257}
\definecolor{MyColor169}{rgb}{0.615812,0.094564,0.61714}
\definecolor{MyColor170}{rgb}{0.610667,0.090204,0.619951}
\definecolor{MyColor171}{rgb}{0.605485,0.085854,0.622686}
\definecolor{MyColor172}{rgb}{0.600266,0.081516,0.625342}
\definecolor{MyColor173}{rgb}{0.595011,0.07719,0.627917}
\definecolor{MyColor174}{rgb}{0.589719,0.072878,0.630408}
\definecolor{MyColor175}{rgb}{0.584391,0.068579,0.632812}
\definecolor{MyColor176}{rgb}{0.579029,0.064296,0.635126}
\definecolor{MyColor177}{rgb}{0.573632,0.060028,0.637349}
\definecolor{MyColor178}{rgb}{0.568201,0.055778,0.639477}
\definecolor{MyColor179}{rgb}{0.562738,0.051545,0.641509}
\definecolor{MyColor180}{rgb}{0.557243,0.047331,0.643443}
\definecolor{MyColor181}{rgb}{0.551715,0.043136,0.645277}
\definecolor{MyColor182}{rgb}{0.546157,0.038954,0.64701}
\definecolor{MyColor183}{rgb}{0.54057,0.03495,0.64864}
\definecolor{MyColor184}{rgb}{0.534952,0.031217,0.650165}
\definecolor{MyColor185}{rgb}{0.529306,0.027747,0.651586}
\definecolor{MyColor186}{rgb}{0.523633,0.024532,0.652901}
\definecolor{MyColor187}{rgb}{0.517933,0.021563,0.654109}
\definecolor{MyColor188}{rgb}{0.512206,0.018833,0.655209}
\definecolor{MyColor189}{rgb}{0.506454,0.016333,0.656202}
\definecolor{MyColor190}{rgb}{0.500678,0.014055,0.657088}
\definecolor{MyColor191}{rgb}{0.494877,0.01199,0.657865}
\definecolor{MyColor192}{rgb}{0.489055,0.010127,0.658534}
\definecolor{MyColor193}{rgb}{0.48321,0.00846,0.659095}
\definecolor{MyColor194}{rgb}{0.477344,0.00698,0.659549}
\definecolor{MyColor195}{rgb}{0.471457,0.005678,0.659897}
\definecolor{MyColor196}{rgb}{0.46555,0.004545,0.660139}
\definecolor{MyColor197}{rgb}{0.459623,0.003574,0.660277}
\definecolor{MyColor198}{rgb}{0.453677,0.002755,0.66031}
\definecolor{MyColor199}{rgb}{0.447714,0.00208,0.66024}
\definecolor{MyColor200}{rgb}{0.441732,0.00154,0.660069}
\definecolor{MyColor201}{rgb}{0.435734,0.001127,0.659797}
\definecolor{MyColor202}{rgb}{0.429719,0.000831,0.659425}
\definecolor{MyColor203}{rgb}{0.423689,0.000646,0.658956}
\definecolor{MyColor204}{rgb}{0.417642,0.000564,0.65839}
\definecolor{MyColor205}{rgb}{0.41158,0.000577,0.65773}
\definecolor{MyColor206}{rgb}{0.405503,0.000678,0.656977}
\definecolor{MyColor207}{rgb}{0.399411,0.000859,0.656133}
\definecolor{MyColor208}{rgb}{0.393304,0.001114,0.655199}
\definecolor{MyColor209}{rgb}{0.387183,0.001434,0.654177}
\definecolor{MyColor210}{rgb}{0.381047,0.001814,0.653068}
\definecolor{MyColor211}{rgb}{0.374897,0.002245,0.651876}
\definecolor{MyColor212}{rgb}{0.368733,0.002724,0.650601}
\definecolor{MyColor213}{rgb}{0.362553,0.003243,0.649245}
\definecolor{MyColor214}{rgb}{0.356359,0.003798,0.64781}
\definecolor{MyColor215}{rgb}{0.35015,0.004382,0.646298}
\definecolor{MyColor216}{rgb}{0.343925,0.004991,0.64471}
\definecolor{MyColor217}{rgb}{0.337683,0.005618,0.643049}
\definecolor{MyColor218}{rgb}{0.331426,0.006261,0.641316}
\definecolor{MyColor219}{rgb}{0.32515,0.006915,0.639512}
\definecolor{MyColor220}{rgb}{0.318856,0.007576,0.63764}
\definecolor{MyColor221}{rgb}{0.312543,0.008239,0.6357}
\definecolor{MyColor222}{rgb}{0.30621,0.008902,0.633694}
\definecolor{MyColor223}{rgb}{0.299855,0.009561,0.631624}
\definecolor{MyColor224}{rgb}{0.293478,0.010213,0.62949}
\definecolor{MyColor225}{rgb}{0.287076,0.010855,0.627295}
\definecolor{MyColor226}{rgb}{0.280648,0.011488,0.625038}
\definecolor{MyColor227}{rgb}{0.274191,0.012109,0.622722}
\definecolor{MyColor228}{rgb}{0.267703,0.012716,0.620346}
\definecolor{MyColor229}{rgb}{0.261183,0.013308,0.617911}
\definecolor{MyColor230}{rgb}{0.254627,0.013882,0.615419}
\definecolor{MyColor231}{rgb}{0.248032,0.014439,0.612868}
\definecolor{MyColor232}{rgb}{0.241396,0.014979,0.610259}
\definecolor{MyColor233}{rgb}{0.234715,0.015502,0.607592}
\definecolor{MyColor234}{rgb}{0.227983,0.016007,0.604867}
\definecolor{MyColor235}{rgb}{0.221197,0.016497,0.602083}
\definecolor{MyColor236}{rgb}{0.21435,0.016973,0.599239}
\definecolor{MyColor237}{rgb}{0.207435,0.017442,0.596333}
\definecolor{MyColor238}{rgb}{0.200445,0.017902,0.593364}
\definecolor{MyColor239}{rgb}{0.193374,0.018354,0.59033}
\definecolor{MyColor240}{rgb}{0.186213,0.018803,0.587228}
\definecolor{MyColor241}{rgb}{0.17895,0.019252,0.584054}
\definecolor{MyColor242}{rgb}{0.171574,0.019706,0.580806}
\definecolor{MyColor243}{rgb}{0.16407,0.020171,0.577478}
\definecolor{MyColor244}{rgb}{0.156421,0.020651,0.574065}
\definecolor{MyColor245}{rgb}{0.148607,0.021154,0.570562}
\definecolor{MyColor246}{rgb}{0.140603,0.021687,0.566959}
\definecolor{MyColor247}{rgb}{0.132381,0.022258,0.56325}
\definecolor{MyColor248}{rgb}{0.123903,0.022878,0.559423}
\definecolor{MyColor249}{rgb}{0.115124,0.023556,0.555468}
\definecolor{MyColor250}{rgb}{0.10598,0.024309,0.551368}
\definecolor{MyColor251}{rgb}{0.096379,0.025165,0.547103}
\definecolor{MyColor252}{rgb}{0.086222,0.026125,0.542658}
\definecolor{MyColor253}{rgb}{0.075353,0.027206,0.538007}
\definecolor{MyColor254}{rgb}{0.063536,0.028426,0.533124}
\definecolor{MyColor255}{rgb}{0.050383,0.029803,0.527975}

\VerbatimFootnotes

\newcommand*\openquote{\makebox(25,-15){\scalebox{5}{\hspace*{-.4cm}\rm{``}}}}
\newcommand*\closequote{\makebox(5,-20){\scalebox{5}{\hspace*{ .2cm}\rm{''}}}}
\definecolor{Azure}{rgb}{0.94, 1.0, 1.0} 
\colorlet{shadecolor}{Azure}
\makeatletter
\newif\if@right
\def\shadequote{\@righttrue\shadequote@i}
\def\shadequote@i{\begin{snugshade}\begin{quote}\openquote}
\def\endshadequote{%
  \if@right\hfill\fi\closequote\end{quote}\end{snugshade}}
\@namedef{shadequote*}{\@rightfalse\shadequote@i}
\@namedef{endshadequote*}{\endshadequote}
\makeatother

\newcommand\hfilll{\hspace{0pt plus 1filll}}

\newcommand{\hbra}{
\hbox to .998\textwidth{\vrule width 0.3mm height 1.8mm depth-0.3mm
                    \leaders\hrule height1.8mm depth-1.5mm\hfill
                    \vrule width 0.3mm height 1.8mm depth-0.3mm}}
\newcommand{\hket}{
\hbox to .998\textwidth{\vrule width0.3mm height1.5mm
                    \leaders\hrule height0.3mm\hfill
                    \vrule width0.3mm height1.5mm}}

\usepackage{soul}

\newcommand{\ie}{{\textit{i.e.}}}
\newcommand{\eg}{{\textit{e.g.}}}
\newcommand{\etal}{{\textit{et al.}}}

\newcommand{\appendixRef}[1]{Appendix~\ref{appendix:#1}}
\newcommand{\figureRef}[1]{Figure~\ref{fig:#1}}
\newcommand{\sectionRef}[1]{Section~\ref{section:#1}}
\newcommand{\tableRef}[1]{Table~\ref{table:#1}}

\newcommand{\ceil}[1]{\left\lceil #1 \right\rceil}
\newcommand{\floor}[1]{\left\lfloor #1 \right\rfloor}

\newcommand{\nbW}{{nbWords}}
\newcommand{\nbD}{{nbDocuments}}
\newcommand{\nbI}{{nbIntegers}}

\newcommand{\gain}{$13\%$}
\newcommand{\mailDatasetOne}{Mail1\xspace}
\newcommand{\mailDatasetTwo}{Mail2\xspace}
\newcommand{\mailDatasetThree}{Mail3\xspace}

\newcommand{\cmix}{cmix\xspace}
\newcommand{\Interp}{Interp\xspace}

\definecolor{coll}{HTML}{000090}

\newcommand{\COMMENT}[1]{}

\usepackage{kbordermatrix} 

\begin{document}

\title{Compressing integer lists with Contextual Arithmetic Trits}

\author{
    Yann~Barsamian\textsuperscript{\small{\Letter}}\\
        \textit{Inria (Paris, France)}\\
        \textit{EPI COSMIQ}\\
        \small{\url{yann.barsamian@eursc.eu}}
    \and
    Andr{\'e} Chailloux\\
        \textit{Inria (Paris, France)}\\
        \textit{EPI COSMIQ}\\
        \small{\url{andre.chailloux@inria.fr}}
}

\maketitle

\begin{abstract}
	
Inverted indexes allow to query large databases without needing to search in
the database at each query. An important line of research is to construct
inverted indexes that require a rather small space usage while still allowing
low timings for compression, decompression, and queries.
In this article, we show how to use trit encoding, combined with contextual
methods for computing inverted indexes. We perform an extensive study of
different variants of these methods and show that our method consistently
outperforms the Binary Interpolative Method --- which is one of the golden
standards in this topic --- with respect to compression size. We apply our
methods to a variety of datasets and make available the source code that
produced the results, together with all our datasets.
\end{abstract}

\newpage
\tableofcontents
\newpage


\section{Introduction}
\label{section:extended_abstract}
\subsection{Context}
When faced with a large database, storing the information contained in its
documents is one of the concerns, but the most prominent one is searching the
database. One of the standard tools to help queries is an \emph{inverted
index}. Each document in the database has an associated identifier, and the
index stores, for each word, the (sorted) list of identifiers whose associated
documents contain this word. The index therefore deals with a large number of
strictly ascending integer lists.

\COMMENT{%
In this article, we focus on the size of the resulting index, \ie, we focus on
the \emph{compression ratio} but for algorithms with a running time comparable
to the Binary Interpolative Method. We present a representation for the
integer lists (transformation into a \emph{trit list}) that consistently
outperforms those methods (up to \gain) on databases ranging from a few MB
(\eg, personal e-mails) to many GB (\eg, large Information Retrieval databases
employed for the World Wide Web).

We make available the source code that produced the results, together with all
our datasets.

\subsection{Formal statement of the problem}%
}
Each integer list we want to compress can be viewed as a \emph{bit vector}
where the bit $0$ means that the word does not appear in the document and the
bit $1$ means that the word appears in the document\footnote{%
	Posting lists is the generalization of this case where we also store the
	number of times a word appears in a document. We do not consider this
	generalization in our work.%
}. Together, those bit vectors form a binary matrix, an example is given in
\figureRef{binary_matrix}.

\begin{figure}[!ht]
	$$
	\mbox{ Words } \overset{\mbox{Document}}{\kbordermatrix{&1&2&3&4&5&6&7&8&9&10&11&12&13&14&15&16\\
			1 & 0 & 0 &  0 & 0 & 0 & 0 & 0 & 0 & 0 & 0 & 0 & 1& 0 & 0 & 0 & 1 \\
			2 & 0 & 1 & 0 & 0 & 0 & 0 & 1 & 1 & 0 & 1 & 1 & 0 & 1 & 0 & 0 & 0 \\
			3 & 0 & 1 & 1 & 1 & 0 & 0 & 0 & 0 & 0 & 0 & 0 & 0 & 0 & 0 & 0 & 0 \\
			4 & 0 & 0 & 0 & 0 & 0 & 0 & 0 & 0 & 0 & 0 & 1 & 0 & 0 & 0 & 0 & 0 \\
			5 & 0 & 0 & 0 & 1 & 1 & 1 & 0 & 0 & 1 & 0 & 0 & 0 & 0 & 1 & 0 & 1 \\
	}}
	$$
	\caption{Integer lists represented as a binary matrix $M$ such that
		$M[i, j] = 1 \Leftrightarrow$ the $i^{\text{th}}$ word appears in the $j^{\text{th}}$ document.}
	\label{fig:binary_matrix}
\end{figure}

Storing this full matrix requires $\nbW \cdot \nbD$ bits, which is $80$~bits in
this example\footnote{This assumes we know the values $\nbW$ and $\nbD$.}.
This would be fine if we had to encode any kind of binary matrix but those
arising from inverse indexes have much more structure. Exploiting this
structure has been the key for deriving more and more successful compression
algorithms. We present below two kinds of structure that are identified and
exploited for most real-life integer lists. 


\paragraph{Sparsity.} The most frequent structure that we see is that inverse
indexes are very sparse. Typically, in the databases we consider, binary
matrices have an average density (the frequency of $1$s) below $0.3\%$. This
means it is more efficient to store the list of documents IDs ({\ie} the
position of $1$s) in which each word appears.
Since these form an increasing sequence for each word, it is often more
efficient to store the gaps between consecutive document IDs and to work on the
corresponding gap arrays. The matrix given in \figureRef{binary_matrix} thus
becomes the arrays in \figureRef{bit_vectors_to_gaps}, where the transformation
between position and gaps is pictured. Most efficient methods work on the
gap arrays, even though one of the best-performing ones, the Binary
Interpolative Method~\cite{MoffatStuiver2000}, works directly on the document
ID lists.

\begin{figure}[!ht]
	$$
	\mbox{ Words }
	{\kbordermatrix{
			& \mbox{Document ID lists}\\
			1	& (12,16) \\
			2	& (2,7,8,10,11,13) \\
			3	& (2,3,4) \\
			4	& (11) \\
			5	& (4,5,6,9,14,16) \\
	}}
	\quad \quad 
	\mbox{ Words }
	{\kbordermatrix{
			& \mbox{Gap lists}\\
			1 & (12,4) \\
			2 & (2,5,1,2,1,2) \\
			3 & (2,1,1) \\
			4 & (11) \\
			5 & (4,1,1,3,5,2) \\
	}}
	$$
	\caption{From bit vectors to document ID lists and gap lists. We took the bit vectors from \figureRef{binary_matrix} and derived the document ID lists and the gap lists.}
	\label{fig:bit_vectors_to_gaps}
\end{figure}

\paragraph{Clustering.}

Inverted indexes have a property which is called clustering: the $1$s appearing
in the gap lists are much closer to each other than what they should be if the
document IDs were uniformly distributed. This depends of course on the ordering
of the documents and can be explained as follows: suppose for example that the
documents are newspaper articles sorted by date. If a tornado happens at a
current date then the word `tornado' will appear in many articles close
together {\ie} close to this date, but will appear rarely at other times. This
clustering effect implies not only that small gaps appear more often but also
that they are close to each other on average. Using this structure is the key
for giving the most efficient compression algorithms. \\

Techniques for efficiently representing those integer lists have to reach a
difficult balance between two main parameters: the size of the inverted index,
and the time needed to query it. For instance, the \cmix compression
algorithm~\cite{CMIX2014} is extremely space efficient but constructing and
querying the inverted index takes a prohibitive amount of time --- 22~hours for
a few MB. Regarding more efficient algorithms, the technique that yielded the
best compression ratio was the \Interp Method~\cite{MoffatStuiver2000} from
1996 and extended in 2000 --- where a few MB are handled in a few seconds. Its
compression ratio has been recently outperformed, on some datasets, by the
Packed+ANS2 method~\cite{MoffatPetri2018}. 


There are two main aspects that we can care about when compressing. First, of
course is the size of the compression: the smaller the size, the better the
compression algorithm is. Another important aspect in practice is the
compression time and the query time.

\subsection{Contributions}

In this article, we present a new compression algorithm for integer lists using
\emph{Contextual Arithmetic Trits}. Our paper may be viewed as the happy
wedding between trits $\{0,1,2\}$--- as exploited by Elias in
1955~\cite[31]{Elias1955} --- and a coding satisfying Shannon's
theorem~\cite{Shannon1948}. We chose arithmetic coding, as also described by
Elias~\cite[61--62]{Abramson1963}, to match as closely as possible the entropy.

The idea of our algorithm is simple, we transform each gap list to a trit list
(where the ``2'' symbol corresponds to the comma between gaps as shown in
\figureRef{gaps_to_trits}). Then we encode these trit lists with an arithmetic
contextual method: the knowledge of the last trits gives a good insight into
the possible values of the next ones. We introduce the notion of hybrid
contexts and perform an detailed analysis of different context models.
We note that previous works on inverted index compression also used contextual
encodings: contextual arithmetic encoding on bit vectors%
~\cite{BooksteinKleinRaita1997}, contextual arithmetic encoding on words%
~\cite{Moffat1989}, contextual asymmetric numeral systems~\cite{Duda2013}
encoding on the result of binary packing blocks of integers%
~\cite{MoffatPetri2018}.

We show that our compression algorithm has better compression rates than both
\Interp and Packed+ANS2. We present gains ranging from a few percent up to
$13\%$. We also present a thorough benchmarking with other compression
algorithms and to do so, we re-implemented many other compression techniques
(around~30). 

\paragraph{Limitations and future work.}
The main limitation of our algorithm is that it was designed solely for
achieving the best compression rates. This means that it is currently less
efficient in terms of time to compress/decompress as well as the time to
perform a query (currently the only possible operation is
compression / decompression of the full index in our algorithm, so performing
queries is prohibitive in time). These features are essential in many
applications. Therefore, our algorithm cannot be directly used in practice
competitively. This leaves however interesting research directions for future
work. The main open question is whether it is possible to use asymmetric
numeral systems~\cite{Duda2013} (ANS) instead of arithmetic encoding and use
block encoding with our contextual trit method in order to achieve an efficient
inverted database while preserving some or most of our compression advantage.
We leave this interesting research direction for future work.

\subsection{Overview of our methods and results}
\label{section:overview}
We briefly present the main ideas of our compression algorithms.

\paragraph{Trit encoding.}
The first idea is to encode elements of the each gap array where the comma
between gaps is encoded with a $2$ symbol. The encoding of each number
({ie} gap) $x$ is its binary representation, minus the leading ``$1$'',
plus a closing ``$2$'', \eg, $19_{10} \rightarrow 10011_{2} \rightarrow 00112$.
The ``$2$'' at the end of each encoded integer serves as a delimiter between
two successive integer encodings and because the binary representation of any
number starts with $1$, we can remove it from the trit list before compression.
\figureRef{gaps_to_trits} illustrates the trit encoding on gap lists. 

\begin{figure}[!ht]
	$$
	\mbox{ Words }
	{\kbordermatrix{
			& \mbox{Gap lists}\\
			1 & (12,4) \\
			2 & (2,5,1,2,1,2) \\
			3 & (2,1,1) \\
			4 & (11) \\
			5 & (4,1,1,3,5,2) \\
	}}
\quad \quad 
{\kbordermatrix{
		& \mbox{Trit lists}\\
		1 & 110021002 \\
		2 & 10210121210212102 \\
		3 & 1021212 \\
		4 & 10112 \\
		5 & 100212121121012102 \\
}}
\quad \quad 
{\kbordermatrix{
		& \mbox{Trit lists without leading $1$s}\\
		1 & 1002002 \\
		2 & 02012202202 \\
		3 & 0222 \\
		4 & 0112 \\
		5 & 002221201202 \\
}}
	$$
	\caption{From gap lists to trit lists}
	\label{fig:gaps_to_trits}
\end{figure}

Now the question is how to encode these trits, knowing that they are not random.
We want to take advantage of the clustering the databases we consider. We
therefore add a contextual method on this trit encoding to get our efficient
compression methods.

\paragraph{Arithmetic coding~\cite[61--62]{Abramson1963} for trits.}
Suppose you have a sequence of $n$ trits with an estimate of the proportion of ``$0$'', ``$1$'' and ``$2$'' symbols denoted respectively $P(0), P(1), P(2)$ with $P(0) + P(1) + P(2) = 1$.
If the trit sequence has $n_i$ symbols $i$ for $i \in \{0,1,2\}$ with $n_0+n_1+n_2 = n$, then a perfect arithmetic encoder will encode your sequence using 

$$\left\lceil n_0|\log_2(P(0))| + n_1|\log_2(P(1))| + n_2|\log_2(P(2))|\right\rceil$$ bits.
This is optimal for random trit sequences with respective symbol frequencies of $0, 1, 2$ equal to $P(0), P(1), P(2)$.
However, our trit encoding of the gap lists are not independent because of the clustering.
We therefore add some context in our arithmetic coding.

\paragraph{Context arithmetic coding for trits.}
In context coding, we don't use absolute probabilities $P(0), P(1), P(2)$ but use instead some probabilities that depend on the previously decoded trits --- the context $C$.
In order not to store too many probabilities, we only use as context the last $\gamma$ encoded trits and use the probabilities $P(i \mid C)$ as our probabilities in our arithmetic coding.
We use an arithmetic encoder with less context (or no context at all for the first trit) to handle the first trits.

For example, if $\gamma = 8$ and if we are encoding the tritlist $201021\mathbf{02022021}t$\dots{} then to encode the trit $t$, out of the 14~trits already encoded so far, the last 8~trits are the context (shown in bold), and we use the probabilities $P(0 \mid 02022021), P(1 \mid 02022021), P(2 \mid 02022021)$ in our arithmetic encoder to encode $t$.
This method requires to store $3 \cdot 3^{\gamma}$ probabilities $P(i \mid C)$
(for each trit $i$ and each context $C$), which implies that we cannot have a
very large depth $\gamma$.

We present several improvements to deal with context arithmetic coding for trits. First, we introduce the notion of \emph{hybrid contexts} which allows to have a deeper depth $\gamma$ by storing only partial information about the context $C$. Then, we study both a static and a dynamic way of estimating the probabilities $P(i \mid C)$. Finally, in the static setting where all these probabilities are stored, we show how reducing the number of precision bits for storing the probabilities decreasing the size of the compressed index. 
 We perform also other optimizations: regarding the ordering of the documents, we use the bisection method, which we also use for concurrent methods for a fair comparison. We also analyze the use of batching in \sectionRef{batching}, {\ie} splitting the words set into batches depending on the density of their associated bit vectors and apply a compression on each batch separately. We also analyze whether sorting the words depending on the density helps or not.



The remainder of this paper is organized as follows: in \sectionRef{related}
we introduce the existing techniques.
In \sectionRef{our_method} we present our compression algorithm in more detail.
In \sectionRef{experiments} we show the datasets and our results.
In \sectionRef{interpretation} we present variants of our compression
algorithm.
\sectionRef{conclusion} summarizes the work and presents some future directions.

\section{Existing compression methods}
\label{section:related}

To compress the integer lists, the literature gives us a lot of different
approaches.
\sectionRef{related_compressing} shows what has been done in a first approach,
to find a suitable encoding of the integer lists.
\sectionRef{related_reordering} gives a rapid insight into another approach
which enhances the clustering property of the integer lists, by reordering the
document IDs.

Last but not least, there are of course a lot of other approaches to
compression, that do not target integer lists. The arithmetic coding, that we
use in this work, is one of the multiple possible general-purpose algorithms
to compress data. We note that the state-of-the-art for general-purpose
compression is the \cmix~\cite{CMIX2014} algorithm, a successor of the PAQ8%
~\cite{KnolldeFreitas2012} algorithm. The interested reader may find in this
latter reference an introduction to arithmetic coding, that is used together
with machine learning in those algorithms, to predict the input characters
based on a static dictionary and on the previous characters read from the file
to compress.

\subsection{Compressing integer lists}
\label{section:related_compressing}

We will suppose that, given a database of $N$ documents, the document IDs are
in $\{ 1, 2, \dots, N\}$. If $N < 2^{32}$, this means that each integer list
can be represented by a sequence of 32-bit unsigned integers. ``Raw'' databases
often use this simple strategy, because it is easy to use~\cite{PISA2014,
IC2014Dataset}. However, as already discussed, in a given database, the gaps
tend to be small. We can thus, instead, use more specialized techniques.

{\bf Universal codes.} Some techniques encode each gap independently. In
this category there are some codes that operate on a bit-by-bit basis, \eg,
Unary, Gamma, Delta~\cite{Elias1975, BentleyYao1976},
Zeta~\cite{BoldiVigna2005}, Golomb, Rice~\cite{Golomb1966,
GallagerVanVoorhis1975} codes. Those codes are presented in detail in, \eg,
\cite{WittenMoffatBell1999}. However, codes that operate on a byte-by-byte
basis are often preferred, because they allow faster decoding:
Variable Byte~\cite{ThielHeaps1972} and its variants, \eg, Varint GB%
~\cite{Dean2009}, Varint-G8IU~\cite{StepanovGangolliRoseErnstOberoi2011}. It is
also possible to operate on different bases, \eg, Variable Nibble%
~\cite[\nopp Section~5]{ScholerWilliamsYiannisZobel2002}, which outputs nibbles.

{\bf Block codes.} Instead of treating the sequence gap by gap, a relatively
recent idea is to treat them by block. In this category, the Simple family
of codes encode a variable-size number of gaps in a fixed-size output (\eg,
$32$ bits for Simple9~\cite{AnhMoffat2005}, $64$ bits for Simple8b%
~\cite{AnhMoffat2010}). The Quantities, Multipliers, and eXtractor (QMX)%
~\cite{Trotman2014} code almost falls in this category, as it outputs most of
the time $128$ bits, and sometimes $256$.

Other types of codes treat a fixed-size number of gaps (\eg, $128$ gaps) and
output a variable-size number of bytes. The most notable encoding of this type
is the Binary Packing, closely related to the Frame-of-Reference (FOR)%
~\cite{GoldsteinRamakrishnanShaft1998}
encoding. It first outputs $b$, the number of bits required to store the
largest of the $128$ gaps in binary notation, then outputs the gaps as $128$
$b$-bits numbers. Sometimes, one gap in a block has a big binary magnitude,
whereas all the others have a small one. This is called an \emph{exception}.
Different authors have come with different solutions to mitigate the cost of
such exceptions, and the resulting encodings are  called Patched
Frame-of-reference (Pfor)~\cite{ZukowskiHemanNesBoncz2006}. Codes in this
category can exhibit high throughput thanks to the use of vectorization, \eg,
SIMD-BP128~\cite{LemireBoytsov2015, JavaFastPFOR2013}.

Another possibility to mitigate the cost of big gaps is to dynamically adapt
the block size: Vector of Split Encoding (VSE)~\cite{SilvestriVenturini2010},
Adaptive Frame-of-reference (AFOR)~\cite{DelbruCampinasTummarello2012,
SIRENe2014}.

Last but not least, it is possible to combine one of these methods with another
compression method, \eg, Packed+ANS2~\cite{MoffatPetri2018}.


{\bf Working directly on the document sequence.} Another possibility is to work
on the sequence of document IDs, instead of working on the gap sequence.
The Quasi-Succinct Indices~\cite{Vigna2013} is a relatively recent
re-engineering of the Elias--Fano~\cite{Elias1972, Fano1971} encoding. This
encoding depends on the maximal possible value encoded, so it is quite natural
to apply the same block techniques that we have seen before: encoding document
sequences by fixed-size chunks is more efficient (\eg, with chunks of $128$
IDs), and dynamically adapting the chunk size offers yet another improvement in
the compression ratio: it is the Partitioned Elias--Fano encoding%
~\cite{OttavianoVenturini2014}.

Last, but not least, comes the Binary Interpolative Encoding (\Interp)%
~\cite{MoffatStuiver2000}. This technique has set the standard for compression
ratio since 1996 (the cited paper from 2000 is an extension of the original
one by the same authors), and has been recently outperformed, on some datasets,
by the Packed+ANS2 method~\cite{MoffatPetri2018}. We note that a variant of
\Interp was published in 2006, that yields better running time but not more
compression~\cite{ChengShannChung2006}.


{\bf Working on the bit vector.} Another possibility is to work on the
\emph{bit vector}, sometimes also called \emph{bitmap}, associated with the
document sequence (as shown in \figureRef{binary_matrix}). Working on bitmaps
is mostly done in the database community, whereas working on integer lists is
mostly done in the information retrieval area, but of course the general idea
stays the same in both areas.

We mention it here for completeness, but also because our method is similar to
what has been applied on bit vectors in the past, but on trit vectors.
Indeed, Bookstein \etal~\cite{BooksteinKleinRaita1997} worked on contextual
methods on bit vectors, and we present, in this article, contextual methods on
trit vectors. This method, as other methods that rely on models, can be
enhanced with batching~\cite{MoffatZobel1992}, see \sectionRef{batching}.

A recent study~\cite{WangLinPapakonstantinouSwanson2017} compared
$21$~compression methods for integer lists, and the interested reader will find
in this study references to methods that work on bit vectors.

\underline{Remark:} as a rule of thumb, most methods that work on bit vectors
take more space than methods that work on document sequences, unless the
sequences are very dense.

\subsection{Reordering the document IDs}
\label{section:related_reordering}

As argued before, if the gaps are small --- or equivalently, if the document
IDs in the integer lists are close to one another --- the compression methods
will usually benefit from it. Another path has hence been taken to reduce the
size of an inverted index: to reorder the document IDs.

The interested reader may find more information about this strategy in%
~\cite[\nopp Table~1]{KaneTompa2014}. In the present paper, we used the
state-of-the-art algorithm of~\cite{Dhulipala2016}, more precisely the
implementation from~\cite{MackenzieMalliaPetriCulpepperSuel2019}.

A recent strategy called Clustered Elias--Fano (CPEF)%
~\cite{PibiriVenturini2017, CPEF2017} divides the full dataset into clusters,
and computes a \emph{reference list} for each cluster. Each document sequence
$S$ is then coded as two sequences: the intersection $I$ between $S$ and the
reference list of its cluster, and of course $S \setminus I$. For clustering,
the authors used the same algorithms as the ones usually used in other
reordering strategies. To encode the resulting sequences, they use Partitioned
Elias--Fano~\cite{OttavianoVenturini2014}. What is novel is that the sequences
to encode are now different. We slightly modified their algorithm to improve
the compression ratio, see \appendixRef{interpolative_details}.

Last but not least, some authors take advantage of the highly repetitive nature
of some document collections, \eg, versioned documents: universal indexes for
Highly Repetitive Document Collections (uiHRDC)%
~\cite{ClaudeFarinaMartinez_PrietoNavarro2016} and Pattern Identification
Sequentially (PIS)~\cite{ZhangTongHuangLiangLiStonesWangLiu2016}. It would be
great to see if those algorithms can be used on our datasets to improve
compression. Indeed, a previous e-mail is usually included when we write an
answer or when we transfer it, so e-mail datasets are no doubt suitable for
such algorithms. Unfortunately, we were only able to test the uiHRDC algorithm,
and it gives less good results, because their algorithm was optimized for
in-memory size and not for output-file size.

\section{Detailed presentation of our method and implementation details}
\label{section:our_method}

The main pseudo-code of our method is detailed in \figureRef{pseudo_code}. In
this section, we will detail the different parts.

\begin{figure}
	\begin{minipage}{\textwidth}
	\begin{footnotesize}
	\begin{tabular}{ll}
	\multicolumn{2}{c}{\hbra}\\
	 0 & {\color{blue} // Initialization}\\
	 1 & Pre-process the integer lists \hfilll \sectionRef{pre_processing}\\
	 2 & Set the model parameters according to global properties \hfilll \sectionRef{parameters}\\
	 3 & \textbf{Foreach} possible context $c \in \mathcal{C}$, \hfilll $\mathcal{C}$ is described in \sectionRef{contexts}\\
	 4 & \hspace{.4cm} \textbf{Foreach} trit $t \in \{ 0, 1, 2 \}$,\\
	 5 & \hspace{.8cm} Initialize a counter $occurrences(c, t)$ at $0$\\
	 6 & {\color{blue} // First pass on the data: collect statistics}\\
	 7 & \textbf{Foreach} gap array $g$,\\
	 8 & \hspace{.4cm} Transform $g$ into a trit list $l$ \hfilll Transformation described in \sectionRef{overview}\\
	 9 & \hspace{.4cm} Initialize a context $c$ to $\emptyset$\\
	10 & \hspace{.4cm} \textbf{Foreach} trit $t$ in $l$\\
	11 & \hspace{.8cm} Increment the counter $occurrences(c, t)$\\
	12 & \hspace{.8cm} Update $c$ according to $t$\\
	13 & {\color{blue} // Write the integer lists lengths, parameters and probabilities in the file}\\
	14 & \textbf{Foreach} gap array $g$,\\
	15 & \hspace{.4cm} Write $|g|$ in the file \hfilll \sectionRef{storing_lengths}\\
	16 & Write the model parameters ($k, w, kInit, \dots$) in the file\\
	17 & \textbf{Foreach} possible context $c$,\\
	18 & \hspace{.4cm} Convert the counters $occurrences(c, 0), occurrences(c, 1)$ and $occurrences(c, 2)$ to probabilities $p(c)$\\
	19 & \hspace{.4cm} Write $p(c)$ in the file \hfilll \sectionRef{storing_probas}\\
	20 & {\color{blue} // Second pass on the data: encode the trit lists}\\
	21 & \textbf{Foreach} gap array $g$,\\
	22 & \hspace{.4cm} Transform $g$ into a trit list $l$\\
	23 & \hspace{.4cm} Initialize a context $c$ to $\emptyset$\\
	24 & \hspace{.4cm} \textbf{Foreach} trit $t$ in $l$\\
	25 & \hspace{.8cm} Arithmetic encoding of $t$ according to $p(c)$\\
	26 & \hspace{.8cm} Update $c$ according to $t$\\
	\multicolumn{2}{c}{\hket}\\
	\end{tabular}
	\end{footnotesize}
	\end{minipage}
	\caption{Pseudo-code for our static arithmetic encoding method.}
	\label{fig:pseudo_code}
\end{figure}

\subsection{Pre-processing}
\label{section:pre_processing}

As explained in \sectionRef{related_reordering}, there is a rich literature
that explains that most compression methods can be used together with a method
that reorders the document IDs. We thus first apply graph bisection on each
dataset~\cite{Dhulipala2016, MackenzieMalliaPetriCulpepperSuel2019}. There has
to be somewhere, stored, a bijection between the actual documents and
$\{ 1, \dots, \nbD \}$. We believe that this reordering do not change the size
of this bijection (this bijection is not accounted for in our results).

For our method that uses adaptive arithmetic coding, there is another
reordering that yields superior compression rates: a reordering on the word
IDs. We store the words by increasing density.

\underline{Remark:} this time, we believe that the cost of storing the
bijection between the actual words and $\{ 1, \dots, \nbW \}$ might be slightly
more --- because standard techniques that apply, \eg, when the words are
sorted alphabetically, are no longer available. We note that if we use a stable
sort, the original ordering can be retrieved without needing to store
additional data --- even though we did not implement it.

\subsection{Choosing the contexts}
\label{section:contexts}
Recall that in context arithmetic coding for trits, we need to store all the probabilities $P(i \mid C)$ for contexts $C$ of length $\gamma$. This limits greatly the depth $\gamma$ we can have. In order to increase the depth, we consider hybrid contexts. 
\paragraph{Hybrid contexts.}
The idea is to store only partial information about the contexts $C$. We write $\gamma = k + w$ for some non-negative integers $k, w$ and consider the following context encoding:
\begin{itemize}
	\setlength\itemsep{-0.2em}
	\item On the last $k$ trits of the context, store for each position whether you have a ``$2$'' symbol encoded by T or another symbol (\ie ``$0$'' or ``$1$'') encoded by N meaning not ``$2$''. There are hence $2^k$ possibilities.
	\item On the $w$ previous trits, store only the total number of ``$2$'' symbols. This is an integer in $\{0,\dots,w\}$.
\end{itemize}

Our contexts $C$ are therefore elements of $\{N, T\}^k \times \{0, \dots, w\}$.
If we take our example, with $\gamma = 8$ and $k = 3, w = 5$, and we are
encoding the tritlist $201021${\color{red}$\underbrace{02022}_w$}{\color{blue}$\underbrace{021}_k$}$t$\dots,
then to encode the trit $t$ we use the probabilities $P(0 \mid NTN,3), P(1 \mid NTN,3), P(2 \mid NTN,3)$
in our arithmetic encoder. This means we have to store $2 \cdot ((w+1)2^k)$
probabilities which can be significantly smaller than $2 \cdot 3^{k+w}$. Our
arithmetic encoding will be less efficient for the same $\gamma$ but we will be
able to take much larger values of $\gamma$ which will make our encoder better
overall. Values for $k, w$ typically vary between $5$ and $12$ depending on the
number of integers in the lists and the variant of the algorithm we use.

We would like to have $\gamma = k + w$ trits of context available, but of
course, this is impossible for the first trits to encode. We thus resort, for
those trits, to smaller contexts. The first trit is encoded with an empty
context, the second trit is encoded with a context of only one trit, the third
trit is encoded with a context of only two trits, and so on, until we reach
$kInit$ trits of context. Then, the context of a trit is the last $kInit$
trits. This way, we encode the first trits in each trit list until we have
encoded $k + w$ trits. Recall that we split $\gamma$ in two parts $k + w$
because having $3^\gamma$ different contexts was not efficient. This explains
why we have yet another parameter $kInit$: fixing $kInit$ to $\gamma - 1$ is
possible but is not efficient, thus we use another, smaller, value.

When using a static arithmetic encoding, we make two passes over the data. In a
first pass, we collect, for each context, the number of trits that have this
context. In the second pass, we will encode those trits according to those
occurrences (transformed into probabilities).

We distinguish two kinds of contexts. We would like to have $\gamma = k + w$
trits of context available, but of course, this is impossible for the first
trits to encode. We thus resort, for those trits, to smaller contexts. The
first trit is encoded with an empty context, the second trit is encoded with
a context of only one trit, the third trit is encoded with a context of only
two trits, and so on, until we reach $kInit$ trits of context. Then, the
context of a trit is the last $kInit$ trits. This way, we encode the first
trits in each trit list until we have encoded $k + w$ trits. Recall that we
split $\gamma$ in two parts $k + w$ because having $3^\gamma$ different
contexts was not efficient. This explains why we have yet another parameter
$kInit$: fixing $kInit$ to $\gamma - 1$ is possible but is not efficient, thus
we use another, smaller, value.

\paragraph{First trits --- merge 0s and 1s}
The first $k+w$ trits of each trit list $l$ are encoded with an initial
arithmetic contextual method --- where the context of the $i$-th trit (we start
counting at $1$) is the $\min(i-1, kInit)$ previous trits, where the trits $0$
and $1$ are undistinguished. There is a number of occurrences to track for the initial contexts equal to:

\[
\displaystyle \sum_{i = 0}^{kInit} 3 \times 2^i = 3 \times \dfrac{1 - 2^{kInit + 1}}{1 - 2}
\]

\paragraph{Rest of the trit list --- merge 0s and 1s}
The rest of the trits are encoded with another contextual arithmetic method ---
where the context of a trit is the $k$ previous trits ($0$ and $1$ are
undistinguished) and the number of $2$ in the $w$ trits before those $k$ trits. There is a number of occurrences to track for the general contexts equal to:

$$3 \times (w + 1) \times 2^k.$$

\subsection{Dynamic vs. static estimation of the probabilities $P(i \mid C)$.}
An important aspect of our method is to determine how we compute our estimates
for $P(i \mid C)$. There are two main ways to compute these estimates:
\begin{enumerate}
	\setlength\itemsep{-0.2em}
	\item \textbf{Static:} when compressing the integer lists, we first read
	all the gap lists to determine these probabilities exactly. In order to
	decompress, we need these probabilities so we have to store them in the
	index.
	\item \textbf{Dynamic:} we start from $P(i \mid C) = \frac{1}{3}$ for each
	$i$ and $C$ --- it is highly probable that better initial estimates are
	possible --- and start our encoder with these probabilities. Each time we
	encode a new trit $i$ with context $C$, we update our probabilities 
	and continue our context arithmetic encoding with these new probabilities.
	This method has the drawback of having less precise estimates for
	$P(i \mid C)$, it has $2$~advantages though:
	\begin{itemize}
		\setlength\itemsep{-0.2em}
		\item The probabilities $P(i \mid C)$ can be reconstructed from scratch
		at the decoding phase, by initializing the decoder again with
		$P(i \mid C) = \frac{1}{3}$ for each $i$ and $C$ and updating the
		probabilities as you decode. This means we don't need to store these
		probabilities which reduces the storage of the compressed integer
		lists.
		\item We don't need to make a first pass on the integer list in order
		to compute these $P(i \mid C)$ which improves the running time of the
		method.
	\end{itemize}
\end{enumerate}

We compared these $2$~methods which are very close in terms of storage, the
dynamic estimation is even usually slightly better.

\subsection{Storing the probabilities}
\label{section:storing_probas}
In the \textbf{static} case, for each context $c$, we now have the number of
times, in the trit lists to encode, where we have a $0$ with this context
$( = occurrences(c, 0))$, where we have a $1$ with this context
$( = occurrences(c, 1))$, and where we have a $2$ with this context
$( = occurrences(c, 2))$. Those three occurrences allow us to compute the
probabilities needed by the encoder:

$$\forall t \in \{ 0, 1, 2 \}, P(t \mid C) = \dfrac{occurrences(c, t)}{\displaystyle \sum_{u = 0}^2 occurrences(c, u)}$$

We must now find a good encoding of those numbers $occurrences(c, t)$. We could
keep things exact and store, for each context, the three values. We found that
it was more effective to let the model be a little bit inexact, but only store
two values. To do so, we normalize those occurrences so that their sum is in
the form $2^b - 1$. We choose a number of bits $b$, then approximate
$P(0 \mid C)$ as the closest fraction with denominator $2^b - 1$, and store the
numerator\footnote{%
    A non-zero occurrence must be represented by a non-zero probability for the
    encoder to work, therefore if $P(0 \mid C) \neq 0$, even if it closer to
    $0$ than to $\dfrac{1}{2^b - 1}$, we must approximate it by
    $\dfrac{1}{2^b - 1}$. The arithmetic encoder we use has the restriction
    that $b \leq 30$, and because we need to be able to encode $3$~non-zero
    probabilities, $b \geq 2$.
}. We do the same for $P(1 \mid C)$, and deduce an approximation of
$P(2 \mid C)$ as $1$ minus the two first approximations. For instance if
$P(0 \mid C) = 8.6\%$, $P(1 \mid C) = 53.1\%$ and $P(2 \mid C) = 38.3\%$, and
if $b = 16$, then we store the $16$-bit number $occ'(c, 0) = 5~636$ representing
$P(0 \mid C) \approx \dfrac{5~636}{65~535}$, we store the $16$-bit number
$occ'(c, 1) = 34~799$ representing $P(1 \mid C) \approx \dfrac{34~799}{65~535}$,
and we deduce (without storing it) $occ'(c, 2) = 25~100$ representing
$P(2 \mid C) \approx \dfrac{25~100}{65~535}$. We therefore encode the three
occurrences on only $32$~bits. The more bits we use to encode those normalized
occurrences, the closest this approximate model will be to the actual data, but
the more space it takes to store the probabilities in the index for decoding.

\underline{Remark}: one could think that $0$'s and $1$'s occur with similar
probabilities in contexts, and would want to go even further in saving space on
the occurences storage by storing only one occurrence per context, say $o$, and
choose as approximations of the occurrences $occ'(c, 0) = occ'(c, 1) = o$ and
$occ'(c, 2) = 2^b - 1 - 2o$. We tried it and it is one of our negative results:
our experiments show that this gives less compression, which means that in
fact, the $0$'s and $1$'s are not balanced in contexts. It is therefore a good
idea to merge $0$'s and $1$'s in contexts but not in probabilities.

\medskip

In the \textbf{dynamic} case, reducing the storage of probabilities is out of
concern, since we simply do not store them in the index. Therefore, we just
keep in memory the dynamic occurrences as they are calculated, while
periodically ``forgetting'' some of the occurrences that have been read so far,
in order to adapt to the latest symbols read~\cite{Gallager1978}. In practice,
occurrences are halved every $N$ steps (see \sectionRef{parameters}), so all
occurrences are guaranteed to be always less than $2N$.

\subsection{Choosing the parameters}
\label{section:parameters}

A crucial point in every method that needs parameters is the choice of those
parameters.

\paragraph{Static Arithmetic Coding.} Our parameters, for the static arithmetic encoding, are:

\begin{itemize}
	\item $k, w, kInit$ for the contexts
	\item the number of bits on which to normalize the occurrences
\end{itemize}

We made experiments on our datasets that showed that (a) the best value for $k$
was growing in accordance to $\nbI$, (b) a value for $w$ close to $k$ was the
best choice, (c) all $kInit$ choices neither too small nor too big with respect
to $k$ were all good choices, and (d) normalizing the occurrences on $8$~bits
was a very good choice.

We thus choose our parameters to be:

$$k, w = k+1, kInit = \ceil{k/3} \text{ and number of bits} = 8,$$

where $k$ is chosen as big as possible without making the storage space of the
model, in bits, going further than $2\%$ of $\nbI$.

\underline{Remark:} for any database with $\nbD$ documents in total, if the
last $\ceil{\nbD} - 1$ trits of context are all different from $2$, then the
next trit will always be a $2$ (because there is no gap bigger than $\nbD$).
We thus tried to see if it was a good ide to make this ``free~2'' available to
the eyes of our encoder, by making sure that $k + w$ would be equal to this
value, in the case where a bigger value for $w$ would have been selected
automatically. In practice, higher values of $w$ lead to better compression
rates. We investigated the use of a further splitting of the context (\eg, by
stating $\gamma = k + w_1 + w_2$ where, as before, in the $k$ last trits of
context we keep all the context, but in the $w_1$ next trits of context we only
count the number of $2$'s, as well as in the $w_2$ next trits of context), but
this line of research did not improve our results. \\


\paragraph{Adaptive Arithmetic Coding.}
Unlike in the static case, the symbol occurrences we are handling are always
estimates of the true ones, based on what has been read so far (the encoding is
done in one pass, thus we have never access to the occurrences of the symbols
in the full message).
First, we provide initial estimates of the probabilities, and then, we
periodically ``forget'' some of the occurrences that have been read so far, in
order to adapt to the latest symbols read~\cite{Gallager1978}.
Last, we must also find good values of the parameters, exactly as in the static
case. When using an adaptive arithmetic encoding, we do not need to store the
model in the index, so the choice of those parameters is not according to a
balance between the precision of the model and its size. Results even suggest
that a greater context does not always lead to a greater compression rate, and
it would be interesting to better understand how to choose good parameters from
a theoretical point of view.


Our parameters, for the adaptive arithmetic encoding, are:

\begin{itemize}\setlength\itemsep{-0.1em}
	\item $k, w, kInit$ for the contexts
	\item the initial estimates of the probabilities $P(i \mid C)$, for each
trit $i$ and each context $C$
	\item In adaptive arithmetic coding, there is a parameter $N$ which is the number of steps in the before dividing by two the occurrences
\end{itemize}

After some experiments, we chose our parameters to be:

$$\left\{\begin{array}{l}w = k =  \max\left(\floor{\dfrac{\log \nbI}{a} + b + 0.5}, 7\right), \text{ where } a = 1.67264 \text{ and } b = -2.24758,\\[10pt]
kInit = \min(2 \times k - 1, 8),\\[10pt]
P(i \mid C) = \dfrac{1}{3} \text{ thus initial occurrences of } 1,\\[10pt]
\text{ and } N = 2^{\min(\max(k, 8), 16)}\end{array},\right.$$

\begin{figure}[!ht]
	\centering\includegraphics{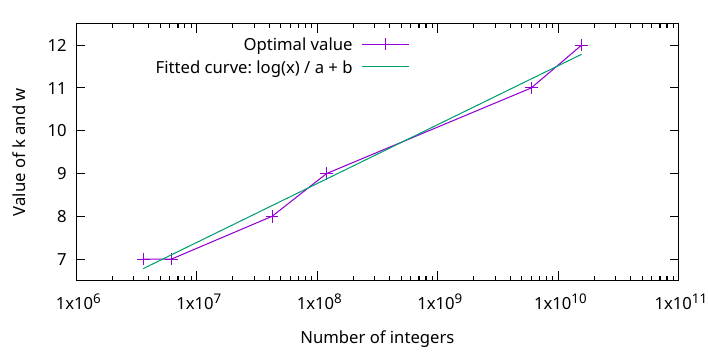}
	\caption{Fitted values of $k$ and $w$ for adaptive contexts.}
    \label{fig:fitted_adaptive_params}
\end{figure}

\underline{Remark:} the values for $a$ and $b$ have been set thanks to a linear
fitting ; we first found optimal values for a subset of our datasets by an
extensive search, and looking at those optimal values suggested this fitting,
see \figureRef{fitted_adaptive_params}.

\subsection{Storing the integer lists lengths}
\label{section:storing_lengths}

We store explicitly, for each integer list, its total length --- the number of
integers in it --- before encoding it.
Indeed, to be able to decode the output of our arithmetic encoder, it suffices
to know in advance when to stop decoding. Here, the number of gaps in an integer
list is equal to the number of ``2'' in its associated trit list. As soon as we
have decoded that number of ``2'', we know we reached the end of the encoding.

Storing those lengths is not mandatory, though, and we picture a variant in
\sectionRef{interpretation}.

We tried different strategies to encode those lengths, and the best performing
one is to encode those lengths with the Delta method.

\underline{Remark:} in our datasets, the encoding of those lengths remains
negligible in front of the encoding of the integer lists. It is possible,
though, that on other datasets, another method would be better to encode those
lengths.

\section{Methodology and results}
\label{section:experiments}

\subsection{Methods studied and implemented}
\label{section:implemented_methods}

Our goal in this paper was to search for the best compression method on
inverted indexes, without any further requirement. There are already a lot of
implementations that can be directly used to compress inverted indexes, but
their purpose is usually different: they also want to minimize the time elapsed
during queries in the index. By doing so, they usually add an overhead.
Needless to say, each team working on the subject has a different way of
dealing with those overheads, and it is most of the time not possible to
compare the compression ratios given in different papers.
Furthermore, it is sometimes useful to combine different techniques, and it
cannot be done without a common framework.
We thus re-implemented methods from the state-of-the-art, in \verb|Java|. The
result is the repository \url{https://gitlab.inria.fr/ybarsami/compress-lists},
that combines the following compression methods:

\begin{itemize}
	\setlength\itemsep{-0.2em}
	\item the methods presented in this paper;
	\item the \Interp method~\cite{MoffatStuiver2000}, together with its
	variants. We rewrote the method with all the optimizations, as
	described in the paper --- note that some optimizations are not applied
	in other papers that use it;
	\item the methods Binary Packing~\cite{GoldsteinRamakrishnanShaft1998},
	FastPFOR~\cite{LemireBoytsov2015}, NewPFD, OptPFD~\cite{YanDingSuel2009},
	Simple9~\cite{AnhMoffat2005}, Simple16~\cite{ZhangLongSuel2008},
	directly taken from the JavaFastPFOR library~\cite{JavaFastPFOR2013};
	\item the methods QMX~\cite{Trotman2014}, Simple8b~\cite{AnhMoffat2010},
	whose implementation straightforwardly derive from the one of Simple9
	from the JavaFastPFOR library in the previous point;
	\item the method AFOR3~\cite{DelbruCampinasTummarello2012}, directly
	taken from the SIRENe library~\cite{SIRENe2014};
	\item the Quasi-Succinct indices~\cite{Vigna2013}, also known as the
	Elias--Fano method, and its refinements with partitioning (PEF%
	~\cite{OttavianoVenturini2014}) and with clustering (CPEF%
	~\cite{PibiriVenturini2017}). The partition algorithm is a simple
	\verb|Java| translation of the code from the original repository%
	~\cite{DS2I2014}. The clustering algorithm has been entirely rewritten
	from the original paper, in order to be able to use not only the
	Elias--Fano method, but other ones as well. We also modified the code
	to add two optimizations, see \appendixRef{interpolative_details};
	\item the Four-State Markov models that work on bit vectors from%
	~\cite{BooksteinKleinRaita1997}. They are contextual arithmetic
	methods, so we used the arithmetic coder from the Reference arithmetic
	coding library~\cite{ArithmeticCoding2011}, and re-implemented the
	ideas of the paper from their description; 
	\item the Bernoulli, Golomb and Rice~\cite{Golomb1966, GallagerVanVoorhis1975}
	methods;
	\item the Unary, Binary, Delta, Gamma~\cite{Elias1975, BentleyYao1976},
	Zeta~\cite{BoldiVigna2005}, Variable Byte~\cite{ThielHeaps1972} and
	Variable Nibble~\cite{ScholerWilliamsYiannisZobel2002} methods, which
	are straightforward to implement given their expressions. It is worth
	noting that for these methods, we explicitly use the fact that we never
	have to code the integer $0$. The encoding is thus the same as
	described in ``Managing Gigabytes''~\cite{WittenMoffatBell1999}, even
	though in other libraries, the authors chose to be able to encode $0$;
	\item last, a variant of the Packed+ANS2 method~\cite{MoffatPetri2018}. We
	re-implemented the algorithm described in the article, and, in our variant,
	we use arithmetic coding instead of ANS to improve the sape usage, hence
	the name Packed+Arithmetic, see \appendixRef{interpolative_details}.
\end{itemize}

There still remains a lot to be said about how to encode an inverted index with a given method.

\subsection{Datasets}

To validate our approach, we used different datasets. Some datasets are
commonly used in the state-of-the-art, and we added e-mails datasets, because
one of our concerns was the use of inverted index to search through e-mails.
Indeed, one of the applications we have in mind, for this work, is an encrypted
mail service. In such a service, the compressed lists would be stored in the
cloud, and would have to be sent to the users that want to use it without using
permanent storage on their device. In that case, unlike most works on inverted
index compression, we need efficient compression even for small datasets.

\begin{itemize}
	\setlength\itemsep{-0.2em}
	\item our personal mails: we computed the inverted indexes of $3$
	personal e-mail boxes, $2$ of which are from the authors.
	\item the ENRON corpus of e-mails~\cite{KlimtYang2004, Enron2015}:
	this corpus contains professional e-mail boxes of $150$ persons. We cleaned the
	original folders by removing duplicate e-mails, as suggested%
	~\cite[\nopp Section~2]{KlimtYang2004}. We present results on the e-mails of
	$4$ users who have the biggest e-mail boxes in the corpus: the e-mails of
	Dasovich, Jones, Kaminski, and Shackleton.
	\item manuals / books: we took four commonly used databases for
	inverted indexes: the Authorized Version (King James') Bible%
	~\cite{Bible1611Dataset}, a bibliographic dataset ``GNUbib'', a collection of
	law documents ``Comact'' (the Commonwealth Acts of Austria), and a collection
	of documents from the Text Retrieval Conference (TREC disks 4 and 5%
	~\cite{TREC1997Dataset}).
	\item web crawling: we took two commonly used databases, which are
	substantially bigger than the other datasets: Gov2 and ClueWeb09%
	~\cite{IC2014Dataset}.
\end{itemize}

We do not index directly all the words in the datasets, but we first normalize
the text: case-folding (``C'' is treated as ``c''), accent removal (``\'e'' is
treated as ``e''), splitting of unicode characters representing two letters
(``\oe'' is treated as ``oe''). The resulting text is then split into words
(which are any sub-string contained into non alphanumerical characters), and
those words are finally stemmed (no stop-words were removed --- frequent words
have integer lists which are the most compressed in indexes).
We used the Porter2 algorithm, improved from the Porter algorithm%
~\cite{Porter1980}, as implemented in the Snowball library~\cite{Snowball2002}.
To index e-mails in MIME format, we made the following choices:

\begin{itemize}
	\setlength\itemsep{-0.2em}
	\item we first parse the e-mails with Apache Mime4j%
	~\cite{ApacheJAMESMime4j2004}
	\item we index the headers (date, subject, from, to, cc, bcc)
	\item we do not index attached files
	\item when we encounter a \verb|text/html| part, we first extract the
	text from the html code with \verb|jsoup|~\cite{Jsoup2009} (to avoid
	indexing words in tags)
	\item when we find a url in the text (``http://...'', ``https://...''
	or ``mailto:...''), we remove what is after question marks (parameters of
	mailto or of php webpages), and remove what is between two consecutive ``/''
	if the number of characters exceeds 30 (to avoid indexing what is usually
	hashed text).
\end{itemize}

While this hand-crafted normalization process is not perfect, it follows the
``Keep It Simple Stupid'' idea.

Tables~\ref{table:datasets1}, \ref{table:datasets2}, \ref{table:datasets3} and
\ref{table:datasets4} give the main properties of those different datasets.


\begin{table}
	\begin{tabularx}{\textwidth}{X*{5}{r}}
		\Xhline{2.5pt}
		Document collection & \mailDatasetOne & \mailDatasetTwo & \mailDatasetThree\\
		\hline
		Number of documents &    32~380 &     7~691 &    26~195 \\
		Number of words     &    90~611 &    48~455 &   116~765 \\
		Number of gaps      & 6~169~953 & 1~394~581 & 4~979~238 \\
		\Xhline{2.5pt}
	\end{tabularx}
	\caption{Main properties of our personal e-mails datasets.}
	\label{table:datasets1}
\end{table}

\begin{table}
	\begin{tabularx}{\textwidth}{X*{5}{r}}
		\Xhline{2.5pt}
		Document collection & Dasovich  & Jones     & Kaminski  & Shackleton & ENRON      \\
		\hline
		Number of documents &    15~748 &    10~646 &    11~348 &   10~576   &    284~982 \\
		Number of words     &    77~871 &    27~691 &    48~174 &   30~250   &    518~412 \\
		Number of gaps      & 3~606~327 & 1~270~340 & 1~701~536 & 1~404~902  & 42~253~227 \\
		\Xhline{2.5pt}
	\end{tabularx}
	\caption{Main properties of datasets from the ENRON corpus.}
	\label{table:datasets2}
\end{table}

\begin{table}
	\begin{tabularx}{\textwidth}{X*{4}{r}}
		\Xhline{2.5pt}
		Document collection &  Bible  & GNUbib    & ComAct     & TREC        \\
		\hline
		Number of documents &  31~102 &    64~267 &    261~829 &     528~155 \\
		Number of words     &   9~423 &    51~910 &    296~351 &   1~098~349 \\
		Number of gaps      & 705~989 & 2~228~876 & 12~919~692 & 119~802~501 \\
		\Xhline{2.5pt}
	\end{tabularx}
	\caption{Main properties of manuals / books datasets.}
	\label{table:datasets3}
\end{table}

\begin{table}
	\begin{tabularx}{\textwidth}{X*{2}{r}}
		\Xhline{2.5pt}
		Document collection & Gov2          & ClueWeb09      \\
		\hline
		Number of documents &    25~205~179 &     50~220~423 \\
		Number of words     &     1~107~205 &      1~000~000 \\
		Number of gaps      & 5~979~715~441 & 15~641~166~521 \\
		\Xhline{2.5pt}
	\end{tabularx}
	\caption{Main properties of web crawling datasets.}
	\label{table:datasets4}
\end{table}

\subsection{Results}
\label{section:results}

We compare here $5$ different methods:
\begin{enumerate}
	\itemsep-0.3em
	\item the (Optimized) \Interp method~\cite{MoffatStuiver2000};
	\item a variant of the \Interp method that works by blocks of 128 integers (to allow better comparison against the next method, that also uses blocking);
	\item a variant of the Packed+ANS2 method~\cite{MoffatPetri2018}, that we call Packed+Arithmetic, that uses arithmetic coding instead of ANS and hence improves compression as explained in \sectionRef{implemented_methods};
	\item our TC Method: TritContext method, with the generic parameters presented in \sectionRef{parameters};
	\item our TCA Method: TritContextAdaptive method, with the generic parameters presented in \sectionRef{parameters}.
\end{enumerate}

Tables~\ref{table:results1}, \ref{table:results2}, \ref{table:results3} and
\ref{table:results4} present our results. Before applying the different
methods, we first apply to the datasets the bisection (b) algorithm%
~\cite{Dhulipala2016, MackenzieMalliaPetriCulpepperSuel2019} and sort (s) the
resulting integer lists by increasing length, hence the (bs) in the dataset
names.
The gain corresponds to the gain in percentage of our best method (in bold)
compared to the \Interp method. All results are written in bits/integer and
include the size required to compute the number of occurrences for each word,
encoded using the Delta~\cite{Elias1975, BentleyYao1976} method. All values are
rounded at $\pm 0.001 $ bits/integer. The gain is rounded $\pm 0.01\%$.

\begin{table}[!ht]
	\begin{tabularx}{\textwidth}{X*{5}{r}}
		\Xhline{2.5pt}
					& \mailDatasetOne(bs) & \mailDatasetTwo(bs) & \mailDatasetThree(bs) \\
		\hline
		\Interp 		& 3.012 & 4.295 & 4.334 \\
		Block-\Interp 		& 3.093 & 4.414 & 4.441 \\
		Packed+Arithmetic	& 2.868 & 4.276 & 4.272 \\
		\hline
		TC   			& 2.692 & 4.097 & 4.114 \\
		TCA 			& \textbf{2.618} & \textbf{3.935} & \textbf{4.026} \\
		\hline
		Gain 			& 13.10\% & 8.39\% & 7.12\% \\
		\Xhline{2.5pt}
	\end{tabularx}
	\caption{Results on our personal e-mails datasets.
    On each dataset, we applied the bisection (b) algorithm%
~\cite{Dhulipala2016, MackenzieMalliaPetriCulpepperSuel2019} and the resulting
integer lists were sorted (s) by increasing length.}
	\label{table:results1}
\end{table}

\begin{table}[!ht]
	\begin{tabularx}{\textwidth}{X*{5}{r}}
		\Xhline{2.5pt}
					& Kaminski(bs) & Jones(bs) & Shackleton(bs) & Dasovich(bs) & ENRON(bs) \\
		\hline
		\Interp 		& 4.520 & 3.843 & 4.073 & 4.038 & 4.043 \\
		Block-\Interp 		& 4.607 & 3.907 & 4.144 & 4.084 & 4.107 \\
		Packed+Arithmetic 	& 4.608 & 3.916 & 4.170 & 4.105 & 4.051 \\
		\hline
		TC 			& 4.440 & 3.695 & 3.979 & 3.945 & 3.868 \\
		TCA 			& \textbf{4.323} & \textbf{3.572} & \textbf{3.890} & \textbf{3.882} & \textbf{3.761} \\
		\hline
		Gain 			& 4.37\% & 7.06\% & 4.49\% & 3.86\% & 6.96\% \\
		\Xhline{2.5pt}
	\end{tabularx}
	\caption{Results on datasets from the ENRON corpus.
    On each dataset, we applied the bisection (b) algorithm%
~\cite{Dhulipala2016, MackenzieMalliaPetriCulpepperSuel2019} and the resulting
integer lists were sorted (s) by increasing length.}
	\label{table:results2}
\end{table}

\begin{table}[!ht]
	\begin{tabularx}{\textwidth}{X*{5}{r}}
		\Xhline{2.5pt}
					& Bible(bs) & GNUbib(bs) & ComAct(bs) & TREC(bs) \\
		\hline
		\Interp 		& \textbf{5.326} & 4.108 & 3.894 & 4.529 \\
		Block-\Interp 		& 5.429 & 4.296 & 4.048 & 4.603 \\
		Packed+Arithmetic	& 5.624 & 4.277 & 4.023 & 4.628 \\
		\hline
		TC   			& 5.358 & 4.037 & 3.824 & 4.552 \\
		TCA   			& 5.354 & \textbf{4.015} & \textbf{3.798} & \textbf{4.489} \\
		\hline
		Gain 			& -0.52\% & 2.25\% & 2.46\% & 0.90\% \\
		\Xhline{2.5pt}
	\end{tabularx}
	\caption{Results on manuals / books datasets.
    On each dataset, we applied the bisection (b) algorithm%
~\cite{Dhulipala2016, MackenzieMalliaPetriCulpepperSuel2019} and the resulting
integer lists were sorted (s) by increasing length.}
	\label{table:results3}
\end{table}

\begin{table}[!ht]
	\begin{tabularx}{\textwidth}{X*{5}{c}}
		\Xhline{2.5pt}
					& Gov2(bs) & Gov2(us) & ClueWeb09(bs) & ClueWeb09(us)\\
		\hline
		\Interp			& 2.690 & 3.485 & 4.309 & 4.999 \\
		Block-\Interp		& 2.722 & 3.522 & 4.338 & 5.028 \\
		Packed+Arithmetic	& 2.708 & 3.462 & 4.274 & 4.982 \\
		\hline
		TC			& 2.579 & 3.344 & 4.123 & 4.803 \\
		TCA			& \textbf{2.575} & \textbf{3.311} & \textbf{4.119} & \textbf{4.763} \\
		\hline
		Gain 			& 4.30\% & 5.00\% & 4.41\% & 4.72\% \\
		\Xhline{2.5pt}
	\end{tabularx}
    \caption{Results on web crawling datasets.
    On each dataset, we applied the bisection (b) algorithm%
~\cite{Dhulipala2016, MackenzieMalliaPetriCulpepperSuel2019} or we let
the documents Ids sorted with a lexicographical sorting on the corresponding
URIs (u)~\cite[\nopp Section~5]{SilvestriVenturini2010},
and the resulting integer lists were sorted (s) by increasing length.}
\label{table:results4}
\end{table}

\newpage
\section{Detailed analysis, interpretation and variants of our method}
\label{section:interpretation}

\subsection{Static contexts}
We first analyze the influence of our main parameters $k$ and $w$ and the way
we store probabilities with static contexts. Throughout this section, we only
store $2$~probabilities per context, see \sectionRef{storing_probas}.

The more $k$ and $w$ grow, the better contexts we have but we have to pay for
the cost of storing the probabilities $P(i \mid C)$, and recall that $C$
contains information about the last $k+w$ trits.

We use one example, the full ENRON dataset (Table \ref{table:datasets2},
rightmost column) and the behavior presented here is very similar for the other
datasets we consider (we include similar tables for the TREC dataset in
\appendixRef{D}). We present two scenarios:

\begin{itemize}
	\item Storing full trit contexts and using the minimum number of bits for
	storing each probability exactly (this minimum number of bits is almost
	always $23$, except for low values of $k$ and $w$ where it is sometimes
	$24$, and even $25$ or $26$ when $k = 0$ and $w \leq 4$).
	\item Merging the symbols ``$0$'' and ``$1$'' (so just put in context
	``$T$'' for the ``$2$'' symbol and ``$N$'' for both symbols ``$0$'' and
	``$1$'') and using $8$~bits for storing each probability.
\end{itemize}

The second scenario has therefore $2$~optimizations that reduce the storage
size of probabilities, at the price of decreasing the accuracy of the model,
therefore increasing the storage size of the compressed trits. We present the
data below. The main number corresponds to size of the compressed lists and we
also specify the space required to store the probabilities (bottom number) and
to store the compressed trits (top number). The size to store the list lengths
is not accounted for in those results. It is of course constant, equal to
$0,207$~MB for the Enron dataset.

Those two scenarios together with the scenario from the next section are
presented in tables where we used the colormap ``plasma''
from~\cite{Smithvan-der-Walt2015} to see at a glance which choices of $k$ and
$w$ lead to the best compression (the same scale of colors is used for the
$3$~tables: colors near yellow --- near white if printed --- indicate better
compressions, while colors near blue --- near grey if printed --- indicate
worse compression).

\begin{figure}[!ht]
	\centering\includegraphics{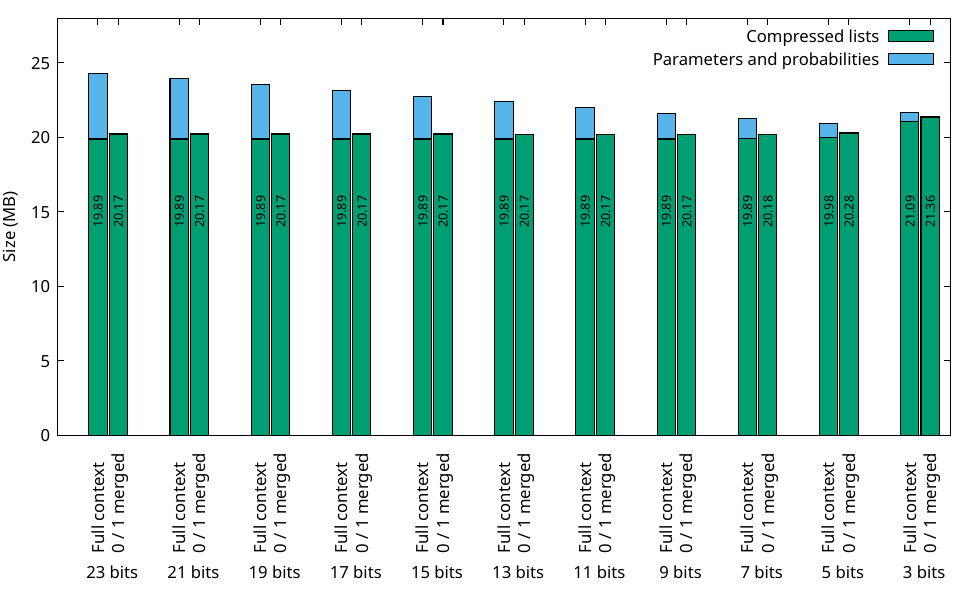}

\caption{Influence of the number of bits used to store probabilities on the
ENRON dataset --- $k = 10$, $w = 12$, $kInit = 4$.}
\label{fig:Enron_k10w12_row_stacked}
\end{figure}

Before having a look at those $2$~detailed tables, we can maybe also have a
glance at \figureRef{Enron_k10w12_row_stacked}, that shows the influence of the
way we store probabilities. As can be seen in this table, the crucial
optimization is to merge ``$0$'' and ``$1$'' in contexts. This means that
instead of storing $2 \cdot ((w+1)3^k)$ probabilities, we store
$2 \cdot ((w+1)2^k)$ probabilities. This of course reduces the precision of the
encoding, but it is more than compensated by the storage saved on probabilities.
The second parameter is the number of bits used to store probabilities. As can
be seen on this graph, this parameter has less influence on the precision of
the encoding, as long as we do not select a too small value for it. We
chose $8$~bits for this parameter, based on our experiments.

\bigskip
\begin{table}[!ht]
\setlength\tabcolsep{1,2pt}
\renewcommand{\arraystretch}{1.5}

\noindent\begin{tabularx}{\linewidth}{|l|*{8}{>{\centering \arraybackslash}X|}} \hline
 & $w = 0$ & $w = 2$ & $w = 4$ & $w = 6$ & $w = 8$ & $w = 10$ & $w = 12$ & $w = 14$ \\\hline
$k = 0$ & 
\cellcolor{MyColor212!50!white}$23.44_{0.00}^{23.44}$ & \cellcolor{MyColor161!50!white}$22.01_{0.00}^{22.01}$ & \cellcolor{MyColor122!50!white}$21.38_{0.00}^{21.38}$ & \cellcolor{MyColor97!50!white}$21.07_{0.01}^{21.06}$ & \cellcolor{MyColor82!50!white}$20.90_{0.01}^{20.89}$ & \cellcolor{MyColor70!50!white}$20.78_{0.01}^{20.77}$ & \cellcolor{MyColor64!50!white}$20.72_{0.01}^{20.71}$ & \cellcolor{MyColor59!50!white}$20.68_{0.01}^{20.67}$\\\hline
$k = 2$ & \cellcolor{MyColor160!50!white}$22.00_{0.00}^{22.00}$ & \cellcolor{MyColor116!50!white}$21.31_{0.00}^{21.30}$ & \cellcolor{MyColor84!50!white}$20.92_{0.01}^{20.91}$ & \cellcolor{MyColor63!50!white}$20.71_{0.01}^{20.70}$ & \cellcolor{MyColor48!50!white}$20.57_{0.01}^{20.56}$ & \cellcolor{MyColor39!50!white}$20.50_{0.01}^{20.49}$ & \cellcolor{MyColor33!50!white}$20.44_{0.01}^{20.43}$ & \cellcolor{MyColor38!50!white}$20.48_{0.01}^{20.47}$\\\hline
$k = 4$ & \cellcolor{MyColor115!50!white}$21.29_{0.00}^{21.28}$ & \cellcolor{MyColor79!50!white}$20.87_{0.01}^{20.86}$ & \cellcolor{MyColor52!50!white}$20.61_{0.01}^{20.60}$ & \cellcolor{MyColor33!50!white}$20.44_{0.01}^{20.43}$ & \cellcolor{MyColor22!50!white}$20.36_{0.01}^{20.34}$ & \cellcolor{MyColor14!50!white}$20.29_{0.01}^{20.28}$ & \cellcolor{MyColor18!50!white}$20.33_{0.01}^{20.31}$ & \cellcolor{MyColor20!50!white}$20.35_{0.02}^{20.33}$\\\hline
$k = 6$ & \cellcolor{MyColor77!50!white}$20.84_{0.01}^{20.83}$ & \cellcolor{MyColor48!50!white}$20.57_{0.02}^{20.55}$ & \cellcolor{MyColor27!50!white}$20.40_{0.03}^{20.37}$ & \cellcolor{MyColor16!50!white}$20.31_{0.04}^{20.27}$ & \cellcolor{MyColor8!50!white}$\mathbf{20.25_{0.05}^{20.20}}$ & \cellcolor{MyColor12!50!white}$20.28_{0.05}^{20.23}$ & \cellcolor{MyColor15!50!white}$20.30_{0.06}^{20.24}$ & \cellcolor{MyColor14!50!white}$20.30_{0.07}^{20.23}$\\\hline
$k = 8$ & \cellcolor{MyColor45!50!white}$20.55_{0.05}^{20.50}$ & \cellcolor{MyColor31!50!white}$20.43_{0.12}^{20.30}$ & \cellcolor{MyColor27!50!white}$20.40_{0.20}^{20.20}$ & \cellcolor{MyColor26!50!white}$20.39_{0.27}^{20.12}$ & \cellcolor{MyColor37!50!white}$20.48_{0.35}^{20.13}$ & \cellcolor{MyColor46!50!white}$20.56_{0.42}^{20.13}$ & \cellcolor{MyColor53!50!white}$20.62_{0.50}^{20.12}$ & \cellcolor{MyColor60!50!white}$20.68_{0.57}^{20.11}$\\\hline
$k = 10$ & \cellcolor{MyColor51!50!white}$20.60_{0.35}^{20.25}$ & \cellcolor{MyColor102!50!white}$21.12_{1.03}^{20.10}$ & \cellcolor{MyColor142!50!white}$21.68_{1.71}^{19.97}$ & \cellcolor{MyColor176!50!white}$22.34_{2.39}^{19.95}$ & \cellcolor{MyColor200!50!white}$22.98_{3.06}^{19.92}$ & \cellcolor{MyColor216!50!white}$23.62_{3.74}^{19.88}$ & \cellcolor{MyColor228!50!white}$24.27_{4.42}^{19.85}$ & \cellcolor{MyColor236!50!white}$24.93_{5.10}^{19.83}$\\\hline
$k = 12$ & \cellcolor{MyColor201!50!white}$23.02_{3.06}^{19.96}$ & \cellcolor{MyColor252!50!white}$28.75_{9.18}^{19.57}$ & \cellcolor{MyColor254!50!white}$34.63_{15.29}^{19.35}$ & \cellcolor{MyColor254!50!white}$40.56_{21.40}^{19.16}$ & \cellcolor{MyColor254!50!white}$46.51_{27.51}^{19.00}$ & \cellcolor{MyColor254!50!white}$52.49_{33.62}^{18.86}$ & \cellcolor{MyColor254!50!white}$58.48_{39.73}^{18.75}$ & \cellcolor{MyColor254!50!white}$64.48_{45.85}^{18.63}$\\\hline
$k = 14$ & \cellcolor{MyColor254!50!white}$46.31_{27.51}^{18.80}$ & \cellcolor{MyColor255!50!white}$99.74_{82.51}^{17.23}$ & \cellcolor{MyColor255!50!white}$153.95_{137.52}^{16.43}$ & \cellcolor{MyColor255!50!white}$208.38_{192.52}^{15.86}$ & \cellcolor{MyColor255!50!white}$262.92_{247.53}^{15.39}$ & \cellcolor{MyColor255!50!white}$317.51_{302.53}^{14.98}$ & \cellcolor{MyColor255!50!white}$372.15_{357.54}^{14.62}$ & \cellcolor{MyColor255!50!white}$426.82_{412.54}^{14.28}$\\\hline
\end{tabularx}
\caption{Size of compressed lists for full ENRON dataset, with distinction between ``$0$'' and ``$1$'' symbols and $23$~bit exact probabilities (except for low values of $k$ and $w$ where it uses sometimes $24$~bits, and even $25$ or $26$ when $k = 0$ and $w \leq 4$). For example, $20.50_{0.01}^{20.49}$ means that the compressed lists use $20.50$~MB, divided into $0.01$~MB for storing the probabilities $P(i \mid C)$ and $20.49$~MB for the rest. All numbers are given $\pm 0.01$~MB.}
\label{table:9}
\end{table}

\COMMENT{%
In \sectionRef{parameters}, we identified automatic parameters to be set with
probabilities normalized on $8$~bits, and such that the total size of the model
would not exceed, in bits, $2\%$ of $\nbI$. We also set our parameters
$w, kInit$ such that $w = k+1$ and $kInit = \ceil{k/3}$.

It is, in the general case, possible to find better parameters. However, of
course, finding the best set of parameters would be very time consuming. For
instance, on Gov2, we found that normalizing probabilities on $16$ bits and
having the total size of the model not exceeding $0.5\%$ of $\nbI$ was
slightly better.%
}

\bigskip 

\begin{table}[!ht]
\setlength\tabcolsep{1,2pt}
\renewcommand{\arraystretch}{1.5}

\noindent\begin{tabularx}{\linewidth}{|l|*{8}{>{\centering \arraybackslash}X|}} \hline
 & $w = 0$ & $w = 2$ & $w = 4$ & $w = 6$ & $w = 8$ & $w = 10$ & $w = 12$ & $w = 14$ \\\hline
$k = 0$ & 
\cellcolor{MyColor212!50!white}$23.44_{0.00}^{23.44}$ & \cellcolor{MyColor161!50!white}$22.01_{0.00}^{22.01}$ & \cellcolor{MyColor123!50!white}$21.39_{0.00}^{21.39}$ & \cellcolor{MyColor99!50!white}$21.09_{0.00}^{21.09}$ & \cellcolor{MyColor84!50!white}$20.92_{0.00}^{20.92}$ & \cellcolor{MyColor73!50!white}$20.81_{0.00}^{20.81}$ & \cellcolor{MyColor70!50!white}$20.78_{0.00}^{20.78}$ & \cellcolor{MyColor65!50!white}$20.73_{0.00}^{20.73}$\\\hline
$k = 2$ & \cellcolor{MyColor161!50!white}$22.00_{0.00}^{22.00}$ & \cellcolor{MyColor118!50!white}$21.32_{0.00}^{21.32}$ & \cellcolor{MyColor86!50!white}$20.94_{0.00}^{20.94}$ & \cellcolor{MyColor66!50!white}$20.74_{0.00}^{20.74}$ & \cellcolor{MyColor51!50!white}$20.60_{0.00}^{20.60}$ & \cellcolor{MyColor47!50!white}$20.56_{0.00}^{20.56}$ & \cellcolor{MyColor40!50!white}$20.50_{0.00}^{20.50}$ & \cellcolor{MyColor44!50!white}$20.54_{0.00}^{20.54}$\\\hline
$k = 4$ & \cellcolor{MyColor117!50!white}$21.31_{0.00}^{21.31}$ & \cellcolor{MyColor82!50!white}$20.90_{0.00}^{20.90}$ & \cellcolor{MyColor56!50!white}$20.65_{0.00}^{20.64}$ & \cellcolor{MyColor37!50!white}$20.48_{0.00}^{20.48}$ & \cellcolor{MyColor31!50!white}$20.43_{0.00}^{20.42}$ & \cellcolor{MyColor22!50!white}$20.36_{0.00}^{20.35}$ & \cellcolor{MyColor25!50!white}$20.39_{0.00}^{20.38}$ & \cellcolor{MyColor27!50!white}$20.40_{0.00}^{20.40}$\\\hline
$k = 6$ & \cellcolor{MyColor81!50!white}$20.89_{0.00}^{20.89}$ & \cellcolor{MyColor53!50!white}$20.62_{0.00}^{20.61}$ & \cellcolor{MyColor32!50!white}$20.44_{0.00}^{20.44}$ & \cellcolor{MyColor23!50!white}$20.37_{0.00}^{20.37}$ & \cellcolor{MyColor13!50!white}$20.29_{0.00}^{20.29}$ & \cellcolor{MyColor16!50!white}$20.31_{0.00}^{20.31}$ & \cellcolor{MyColor17!50!white}$20.32_{0.00}^{20.32}$ & \cellcolor{MyColor15!50!white}$20.31_{0.00}^{20.30}$\\\hline
$k = 8$ & \cellcolor{MyColor52!50!white}$20.61_{0.00}^{20.61}$ & \cellcolor{MyColor29!50!white}$20.41_{0.00}^{20.41}$ & \cellcolor{MyColor19!50!white}$20.33_{0.00}^{20.33}$ & \cellcolor{MyColor7!50!white}$20.24_{0.01}^{20.24}$ & \cellcolor{MyColor8!50!white}$20.25_{0.01}^{20.25}$ & \cellcolor{MyColor8!50!white}$20.25_{0.01}^{20.25}$ & \cellcolor{MyColor7!50!white}$20.24_{0.01}^{20.23}$ & \cellcolor{MyColor5!50!white}$20.23_{0.01}^{20.22}$\\\hline
$k = 10$ & \cellcolor{MyColor28!50!white}$20.41_{0.00}^{20.40}$ & \cellcolor{MyColor16!50!white}$20.31_{0.01}^{20.30}$ & \cellcolor{MyColor3!50!white}$20.21_{0.01}^{20.20}$ & \cellcolor{MyColor4!50!white}$20.22_{0.02}^{20.21}$ & \cellcolor{MyColor3!50!white}$20.22_{0.02}^{20.20}$ & \cellcolor{MyColor1!50!white}$20.20_{0.02}^{20.18}$ & \cellcolor{MyColor0!50!white}$20.19_{0.03}^{20.17}$ & \cellcolor{MyColor0!50!white}$\mathbf{20.19_{0.03}^{20.16}}$\\\hline
$k = 12$ & \cellcolor{MyColor15!50!white}$20.31_{0.01}^{20.30}$ & \cellcolor{MyColor1!50!white}$20.20_{0.03}^{20.18}$ & \cellcolor{MyColor2!50!white}$20.21_{0.04}^{20.17}$ & \cellcolor{MyColor2!50!white}$20.21_{0.06}^{20.15}$ & \cellcolor{MyColor1!50!white}$20.20_{0.08}^{20.13}$ & \cellcolor{MyColor1!50!white}$20.20_{0.09}^{20.11}$ & \cellcolor{MyColor1!50!white}$20.20_{0.11}^{20.09}$ & \cellcolor{MyColor2!50!white}$20.21_{0.12}^{20.08}$\\\hline
$k = 14$ & \cellcolor{MyColor2!50!white}$20.20_{0.03}^{20.17}$ & \cellcolor{MyColor5!50!white}$20.23_{0.10}^{20.13}$ & \cellcolor{MyColor8!50!white}$20.25_{0.17}^{20.09}$ & \cellcolor{MyColor12!50!white}$20.28_{0.23}^{20.05}$ & \cellcolor{MyColor17!50!white}$20.32_{0.30}^{20.02}$ & \cellcolor{MyColor23!50!white}$20.36_{0.36}^{20.00}$ & \cellcolor{MyColor29!50!white}$20.41_{0.43}^{19.99}$ & \cellcolor{MyColor36!50!white}$20.47_{0.49}^{19.98}$\\\hline
\end{tabularx}
\caption{Size of compressed lists for full ENRON dataset, without distinction between ``$0$'' and ``$1$'' symbols and $8$~bit probabilities. For example, $20.20_{0.09}^{20.11}$ means that the compressed lists use $20.20$~MB, divided into $0.09$~MB for storing the probabilities $P(i \mid C)$ and $20.11$~MB for the rest. All numbers are given $\pm 0.01$~MB. The optimal value is outside of the table: a size of $20.18$~MB can be reached for $k = 10$ and $w = 21$.}
\label{table:10}
\end{table}

We have a few takeaways when looking at these two tables and this histogram:
\begin{itemize}\setlength\itemsep{-0.2em}
	\item Without merging $0$ and $1$s, we can see that a few positions of context can help but then, the space required to store the probabilities becomes prohibitive
	\item In our optimized compression scheme, we have a slightly smaller size of our compressed database.
	\item Another important feature is that with $8$~bit probabilities and with merged $0$ and $1$s, the size of the compressed lists is much more stable w.r.t. the parameters $k$ and $w$. This is what allows to have a predetermined set of $k,w,kInit$ (as a function only of the number of integers of the document) which is close to the optimal.
	\item When comparing with other data, using $8$~bits for probabilities helps a little bit but the most important optimization is the merging of ``$0$'' and ``$1$'' in the context.
\end{itemize}

\subsection{Adaptive contexts}

We also present how the $k$ and $w$ parameters influence the adaptive context method. Here, we don't need to store the probabilities $P(i \mid C)$ since they are constructed as the compression is performed and are calculated as a part of the decompression step. 
We present below a table of values for the ENRON dataset for this method. 

\bigskip

\begin{table}[!ht]

\setlength\tabcolsep{1.5pt}
\renewcommand{\arraystretch}{1.5}

\noindent\begin{tabularx}{\linewidth}{|l|*{8}{>{\centering \arraybackslash}X|}} \hline
 & $w = 0$ & $w = 2$ & $w = 4$ & $w = 6$ & $w = 8$ & $w = 10$ & $w = 12$ & $w = 14$ \\\hline
$k = 0$ & 
\cellcolor{MyColor111!50!white}$21.23$ & \cellcolor{MyColor64!50!white}$20.72$ & \cellcolor{MyColor38!50!white}$20.49$ & \cellcolor{MyColor28!50!white}$20.40$ & \cellcolor{MyColor22!50!white}$20.36$ & \cellcolor{MyColor22!50!white}$20.35$ & \cellcolor{MyColor21!50!white}$20.35$ & \cellcolor{MyColor17!50!white}$20.32$\\\hline
$k = 2$ & \cellcolor{MyColor69!50!white}$20.76$ & \cellcolor{MyColor35!50!white}$20.47$ & \cellcolor{MyColor11!50!white}$20.27$ & \cellcolor{MyColor0!50!white}$20.17$ & \cellcolor{MyColor0!50!white}$20.09$ & \cellcolor{MyColor0!50!white}$20.04$ & \cellcolor{MyColor0!50!white}$20.00$ & \cellcolor{MyColor0!50!white}$19.99$\\\hline
$k = 4$ & \cellcolor{MyColor39!50!white}$20.50$ & \cellcolor{MyColor8!50!white}$20.25$ & \cellcolor{MyColor0!50!white}$20.08$ & \cellcolor{MyColor0!50!white}$19.97$ & \cellcolor{MyColor0!50!white}$19.89$ & \cellcolor{MyColor0!50!white}$19.81$ & \cellcolor{MyColor0!50!white}$19.80$ & \cellcolor{MyColor0!50!white}$19.84$\\\hline
$k = 6$ & \cellcolor{MyColor4!50!white}$20.22$ & \cellcolor{MyColor0!50!white}$20.03$ & \cellcolor{MyColor0!50!white}$19.90$ & \cellcolor{MyColor0!50!white}$19.80$ & \cellcolor{MyColor0!50!white}$19.73$ & \cellcolor{MyColor0!50!white}$19.71$ & \cellcolor{MyColor0!50!white}$19.77$ & \cellcolor{MyColor0!50!white}$19.85$\\\hline
$k = 8$ & \cellcolor{MyColor0!50!white}$19.98$ & \cellcolor{MyColor0!50!white}$19.85$ & \cellcolor{MyColor0!50!white}$19.76$ & \cellcolor{MyColor0!50!white}$19.69$ & \cellcolor{MyColor0!50!white}$\mathbf{19.66}$ & \cellcolor{MyColor0!50!white}$19.73$ & \cellcolor{MyColor0!50!white}$19.82$ & \cellcolor{MyColor0!50!white}$19.85$\\\hline
$k = 10$ & \cellcolor{MyColor0!50!white}$20.00$ & \cellcolor{MyColor0!50!white}$19.98$ & \cellcolor{MyColor0!50!white}$19.92$ & \cellcolor{MyColor0!50!white}$19.88$ & \cellcolor{MyColor0!50!white}$19.87$ & \cellcolor{MyColor0!50!white}$19.88$ & \cellcolor{MyColor0!50!white}$19.92$ & \cellcolor{MyColor0!50!white}$19.94$\\\hline
$k = 12$ & \cellcolor{MyColor0!50!white}$20.18$ & \cellcolor{MyColor0!50!white}$20.18$ & \cellcolor{MyColor0!50!white}$20.15$ & \cellcolor{MyColor0!50!white}$20.17$ & \cellcolor{MyColor0!50!white}$20.19$ & \cellcolor{MyColor4!50!white}$20.22$ & \cellcolor{MyColor9!50!white}$20.26$ & \cellcolor{MyColor16!50!white}$20.31$\\\hline
$k = 14$ & \cellcolor{MyColor36!50!white}$20.47$ & \cellcolor{MyColor39!50!white}$20.49$ & \cellcolor{MyColor39!50!white}$20.50$ & \cellcolor{MyColor41!50!white}$20.52$ & \cellcolor{MyColor45!50!white}$20.54$ & \cellcolor{MyColor48!50!white}$20.57$ & \cellcolor{MyColor51!50!white}$20.60$ & \cellcolor{MyColor54!50!white}$20.63$\\\hline
\end{tabularx}
\caption{Size of compressed lists for full ENRON dataset with adaptive contexts. 
All numbers are given $\pm 0.01$~MB.}
\end{table}

We see again that there is a sweet spot in the choice of $k$ and $w$.
Interestingly, when taking large contexts, the accumulated information is
somehow less relevant and gives less compression.

\subsection{Time results}
\label{section:timings}

\tableRef{timings} shows the timings of our methods compared to others. While
writing our \verb|Java| implementations, we did not try to optimize any part
of the code. If one is interested into optimizing the timings, we are pretty
sure that the implementation we used for the arithmetical encoding can be
improved, but others could probably be improved also.

Experimental hardware is an Intel Core i5-6300U CPU @ 2.40GHz (Skylake-U), with
16~GB of RAM. All reported numbers are computed as the mean of $10$~independent
results.

Without any optimization, when using contextual trits, the compression takes up
to $6.1$~times as much time as when using \Interp, and up to $8.0$~times as much
time as when using delta.
When using the adaptive contextual trits, the compression takes up to
$4.8$~times as much time as when using \Interp, and up to $6.3$~times as much
time as when using delta.

Without any optimization, when using contextual trits, the decompression takes
up to $1.6$~times as much time as when using \Interp, and up to $1.4$~times as
much time as when using delta.
When using the adaptive contextual trits, the decompression takes up to
$1.6$~times as much time as when using \Interp, and up to $1.5$~times as much
time as when using delta.

\begin{table}[!ht]
\noindent\begin{tabularx}{\linewidth}{l*{8}{>{\centering \arraybackslash}X}}
    \Xhline{2.5pt}
        & \multicolumn{2}{c}{Bible(bs)} & \multicolumn{2}{c}{Mail1 (bs)} & \multicolumn{2}{c}{ENRON(bs)} & \multicolumn{2}{c}{TREC(bs)}\\
        & Comp & Decomp & Comp & Decomp & Comp & Decomp & Comp & Decomp\\
    \Xhline{2.5pt}
Interpolative & 0.43 & 1.60 & 0.15 & 0.82 & 0.12 & 0.98 & 0.12 & 1.06 \\
Delta & 0.18 & 1.57 & 0.07 & 0.84 & 0.08 & 1.03 & 0.09 & 1.21 \\
Contextual trits & 0.90 & 2.20 & 0.44 & 1.13 & 0.60 & 1.45 & 0.74 & 1.69 \\
Adaptive contextual trits & 0.83 & 2.35 & 0.39 & 1.16 & 0.48 & 1.46 & 0.58 & 1.72 \\
    \Xhline{2.5pt}
\end{tabularx}

\caption{Time to compress or decompress the integer lists from and to .csv file (in $\mu$s per integer) for different methods on various datasets.}
\label{table:timings}
\end{table}

\underline{Remark:} even though we did not investigate the optimization of
those timings, one can note that using arithmetic encoding on trit lists is way
faster than using this encoding on bit vectors, as there are far less trits
than bits to encode. For instance, the Enron dataset contains $147~738~088~584$~bits
but only $118~861~983$~trits; the TREC dataset contains $580~098~516~095$~bits
but only $401~422~437$~trits (around $1000$~times less in each case). As a
consequence and to give a point of comparison, the Bookstein method that uses
arithmetic encoding on the bit vector uses around $1$~h of computation time to
compress the integer lists associated with the Enron dataset, instead of $25$~s
for our contextual trits.

If one wants to optimize further the timings required to compress trits, it is
probably possible to use asymmetric numeral systems~\cite{Duda2013} as they
provide a compression close to arithmetic encoding, with far superior speed.

\subsection{Other variants}
There are many other ways of performing contexts which, in our settings, gave equivalent or slightly worst results. We still think it's interesting to include some of these ideas that could give improvements in other contexts. These are included in \appendixRef{C}.

\section{Conclusion}
\label{section:conclusion}

In this article, we developed a novel idea to compress inverted indexes, that
has better compression rates than the \Interp and Packed+ANS2 methods, two
methods that are the state-of-the-art regarding compression ratios.
We applied our method to a wide range of datasets, including e-mail datasets.
In many applications, the most useful feature is not necessarily the
compression rate and we hope that, in the future, the trits we used in this
paper can be used together with other algorithms, to provide better time
complexities, as well as performing local queries with a minimal loss 

We mentioned contributions taking advantage of the highly repetitive nature of
some document collections, \eg, versioned documents. We hope that, in the
future, it will be possible to combine those ideas with ours.

Last but not least, we think that our methods can still be enhanced. Throughout
this paper, we highlighted that the contextual probabilities of the trits are
the key to the success of our methods. We tried, in different ways, to reach a
balance between the precision of the probabilities (bigger contexts) and the
cost of the model (in case it needs to be stored) or the time needed to read
sufficiently enough trits to have a good estimate of the true probabilities (in
case the model is built on-the-fly). What we did not do is to look at what
those probabilities look like. We think that it could probably enhance our
methods if we could look at the probabilities and extract a pattern from them,
or use machine learning to enhance prediction, as is done in \cmix%
~\cite{CMIX2014} and PAQ8~\cite{KnolldeFreitas2012}. This is left for future
work.

\vspace{0.5em}
\noindent {\bf Acknowledgments.}
We would like to thank A.~Moffat for unearthing the GNUbib and ComAct datasets.
We would like to thank K.~Merckx for porting to our attention the ENRON
dataset.
We would like to thank the authors of the various softwares we used in this
experiment. No comparison could have been done without their previous efforts.
We directly included in our code, by alphabetical order,
``JavaFastPFOR''~\cite{JavaFastPFOR2013},
``jsoup''~\cite{Jsoup2009},
``Mime4j''~\cite{ApacheJAMESMime4j2004},
``Reference arithmetic coding''~\cite{ArithmeticCoding2011},
``SIRENe''~\cite{SIRENe2014},
and ``Snowball''~\cite{Snowball2002}.
We also used the following softwares, by alphabetical order,
``CPEF''~\cite{CPEF2017} (to split the dataset by clusters),
``Matplotlib''~\cite{Smithvan-der-Walt2015} (to have a black-and-white friendly
colormap for our result tables), and
``PISA''~\cite{PISA2014} (to reorder the datasets using graph bisection).
Following their path, we also provide the source code of our experiments, see
\url{https://gitlab.inria.fr/ybarsami/compress-lists}.
Special thanks to the Thalys train which made this collaboration easy.
Last but not least, we would like to thank our anonymous reviewers for their
comments. Everybody writes that, but in our case, their insights and questions
made us improve substantially the quality of our paper.

\newpage

\appendix

\section{Detailed comparison to the state-of-the-art}
\label{appendix:interpolative_details}

Comparing implementations is always very difficult and error-prone, even when
one is devoting the best they can. Before going to the tables
\tableRef{interpolative_implems}, \ref{table:interpolative_implems_filtered} and
\ref{table:interpolative_implems_unfiltered}, we would like to say a few words.

First, comparing on the exact same datasets is hard. We were able to download
the datasets Gov2 and ClueWeb09 from~\cite{IC2014Dataset}, which were already
filtered from the originals. Other papers use a different filtering,
sometimes no filtering at all is used\dots{} and of course results differ
depending on the filtering used, see for instance~\cite[\nopp Table 24 (page 115)]{Pibiri2018}
and~\cite[\nopp Section~4.1, Table~II]{PibiriVenturini2017}.
The Gov2 and ClueWeb09 datasets we have at our disposal are already versions of
the datasets where the small integer lists were removed (in Gov2 (resp.
ClueWeb09), the integer lists have lengths no less than $100$ (resp. $250$)
integers). These datasets are thus really close to the datasets filtered in
\cite{MoffatPetri2018}. To reproduce as closely as possible the work in
\cite{PibiriVenturini2017}, we also present results, in this appendix, on
datasets where even more small integer lists are removed (in Gov2(*f) (resp.
ClueWeb09(*f)), the integer lists have lengths no less than $14182$ (resp.
$16618$), to match~\cite[\nopp Section~4.1, Table~II]{PibiriVenturini2017}),
see the properties of those modified datasets in \tableRef{datasets5}.
It would be interesting to always take all the integer lists, and explain the
strategies on the different sub-lists. In this regard, our implementation
provides a ``Meta'' method which is able to use different compression
algorithms on different integer lists.

\begin{table}
	\begin{tabularx}{\textwidth}{X*{4}{r}}
		\Xhline{2.5pt}
		Document collection & \multicolumn{2}{c}{Gov2} & \multicolumn{2}{c}{ClueWeb09} \\
		                    & Original      & Filtered      & Original       & Filtered \\
		\hline
		Number of documents &    25~205~179 &    25~205~179 &     50~220~423 &     50~220~423 \\
		Number of words     &     1~107~205 &        25~353 &      1~000~000 &         51~663 \\
		Number of gaps      & 5~979~715~441 & 5~272~947~888 & 15~641~166~521 & 14~261~647~820 \\
		\Xhline{2.5pt}
	\end{tabularx}
	\caption{Main properties of web crawling datasets, with and without filtering.}
	\label{table:datasets5}
\end{table}

%
%
%
%
%
%
%

\begin{figure}[!ht]
\begin{verbatim}
int[] documentArray = new int[] {
        84, 85, 510, 941, 946, 965, 978, 1008, 1009, 1774, 1862, 2248, 2254,
        2755, 2756, 3494, 3495, 3716, 4428, 4462, 4676, 5218, 5219, 5430, 5455,
        5470, 6007, 6229, 6408, 6467, 6500, 6601, 6654, 6850, 7757, 8261, 8262,
        8263, 8264, 8265, 8324, 8359, 8423, 8438, 8808, 9413, 9739, 9885, 10512,
        10766, 10842, 10962, 11124, 11140, 11141, 11188, 11222, 11780, 12146,
        12148, 12415, 12455, 12456, 12644, 12736, 13643, 14131, 14153, 14172,
        14239, 14240, 14250, 14254, 14262, 14596, 14860, 15032, 15033, 15042,
        15043, 15428
};
\end{verbatim}

	\centering\includegraphics{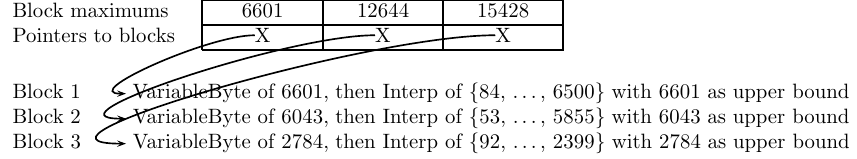}

\justifying$\{84, \dots, 6500\}$ are the first $31$ document IDs.
$\{53, \dots, 5855\}$ are the document IDs number $33$ to $63$ from which
$6601$ was subtracted.
$\{92, \dots, 2399\}$ are the document IDs number $65$ to $80$ from which
$12644$ was subtracted.
\caption{Example encoding of a document array of length $81$ with blocks of
size $32$ with a non-optimized Block-\Interp method as found in the literature.}
\label{fig:document_array}
\end{figure}

Second, not all papers have the same objectives in mind. Some focus on the time
needed to compress / decompress (\eg, using vectorization), on the compression
size, on the RAM usage, on the time to query the database\dots{}
In our paper, we chose to focus solely on the compression size, and as we saw
in \sectionRef{timings}, this means that while we achieve a better compression
rate, this is at the cost of a greater compression time, and at the cost of not
supporting the same set of query operations than other methods.
Therefore, it would not be fair from us to compare our results directly to
other implementations that optimize on a wider range of criteria. This explains
why we chose to re-implement most methods (in addition to instrumenting other
implementations): to compare always to the best possible compression rates of
other methods while forgetting other criteria.
Papers that fall into this category usually decompose large gap lists in
smaller blocks. Of course, as far as queries are concerned, authors are
interested not only on the compressed size of the index, but also on the time
to query it. With this in mind, it is obvious that on large gap arrays, it is
faster to decompress only a block rather than the full array.
This explains why those papers compare their results not to the original
\Interp method, but to a modified version of the \Interp method, where an
integer list is not coded as a whole, but rather divided by blocks (\eg, of
size $128$), with additional memory overhead to enable fast queries inside
those blocks.
To illustrate this fact, let us look at the document array in \figureRef{document_array},
together with the encoding we found in other implementations.
We see from this figure that the VariableByte encodings of $6601$, $6043$ and
$2784$ can be avoided, as they can be recomputed from the block maximums array.
This optimization leads to a better space usage and we believe that it would
not use more time (neither to construct, nor to query) --- it could even
marginally use less time because a subtraction of two integers already in the
memory is probably quicker than a decoding of the VariableByte method.
When comparing to their implementations, we therefore instrumented their code
to remove those additional VariableByte encodings.
Wasting a word per block of $128$~integers makes a noticeable difference on the
resulting index size, hence, we think that a comparison to an optimized
implementation of the block-\Interp is needed, with and without enforcing
fast-skipping of blocks. This comparison is done in this appendix.

In our repository\footnote{%
\url{https://gitlab.inria.fr/ybarsami/compress-lists}%
}, we first re-implemented the different versions of
\Interp. Then, we re-implemented the cluster method, with two changes
from the original CPEF method: (a) we only considered the frequency-based
construction for the reference list, to be able to also use the cluster method
in concordance with other methods than the PEF method and (b) we found and
implemented an optimization with respect to index size called ``shrinking''.
Recall that each document sequence $S$ is coded as two sequences: the
intersection $I$ between $S$ and the reference list of its cluster, and of
course $S \setminus I$. For our optimization, instead of coding directly the
integers in $S \setminus I$, we first apply a mapping $m$ from $S \setminus I$
to $\mathbb{N}$ where $m(x) = x - \#\{i \in \mathrm{reference~list} \mid i < x\}$.
Last but not least, while reimplementing the Packed+ANS2 method, we used
arithmetic encoding which gives slighlty better compression ratio. In our
implementation, (a) we did not optimize the RAM usage
(hence, we do not use ``ANSmsb'' from \cite[\nopp Section~3.3]{MoffatPetri2018})
but (b) we keep all the	$16 \times 17/2 = 136$ contexts instead of merging them
to have only $64$. The use of $64$~contexts in the original paper was done in
order to keep the number of bits used for the context selector as low as $6$
(instead of $8$), thus gaining $2$~bits per block. We found that encoding the
selector with an arithmetic coder takes less than $6$~bits, even with the
$136$~contexts\footnote{%
    The selector uses between $4.5$ and $4.9$~bits on average for mail datasets,
    and $4.6$~bits for the Gov2 dataset.},
and permits to distinguish more contexts, thus gaining space also on the
compression of the gaps.

Having said that, we can now show the results of these different methods in
\tableRef{interpolative_implems}.
As can be seen in the table, the \Interp method is almost always the best, even
though the Packed+Arithmetic method is close --- sometimes better.
Blocking makes a noticeable difference ($2.7\%$ on \mailDatasetOne, $1.6\%$ on TREC,
$1.2\%$ on Gov2, more if we add padding), and we recall that in the current
article, we provide comparisons against the original \Interp method, which
provides the best compression ratio.
We can also see that clusters help more on the original URIs-sorted datasets,
but are sometimes detrimental on the bisection-sorted datasets, depending on
the method with which they are associated.
Last but not least, those tables provide results only on the gap lists, and not
on the full posting lists as in other papers. By comparing the results we give
to the results given in other papers, it is possible to think that, with
respect to space usage, results on the ``docs'' are not always similar to
results on the ``freqs''. It would be interesting to see if it possible to use
different methods on those two parts of the postings lists. Indeed, those
files do not have the same kind of structure as for the distribution of the
integers contained in them.

We provide below tables presenting our data.
\begin{itemize}
	\item In Table~\ref{table:interpolative_implems}, we compare the total index size of our method and other methods for three different datasets. The Block-* methods use blocks of fixed-size $128$. The PEF* methods use
	dynamic block sizes, where the block sizes are computed to minimize the total
	cost (opt means that the optimum is found thanks to dynamic programming, approx
	means it is an $(1 + \epsilon)$ approximation of that optimum, thanks to the
	algorithm from the original article~\cite{OttavianoVenturini2014}).
	For all block methods, ``padding'' means that each block encoding is padded to
	the end of one byte, whereas ``no padding'' means that there is no such
	padding.
	The ``opt'' algorithm takes quadratic time, and thus did not finish in a
	reasonable amount of time on bigger datasets. This is why some results are
	missing. On each dataset, we applied the bisection (b) algorithm%
	~\cite{Dhulipala2016, MackenzieMalliaPetriCulpepperSuel2019}.
\item In Table~\ref{table:interpolative_implems_filtered}, we provide similar data on filtered web crawling datasets. The Gov2 (resp. ClueWeb09) dataset from \cite{IC2014Dataset} has been filtered
(f) by removing all integer lists whose size was strictly less than $14~182$
(resp. $16~618$), see \cite[\nopp Table~II]{PibiriVenturini2017}.

In our implementation of the CPEF method, we use the frequency-based method to
compute the reference lists. For a reference list size of 800K for Gov2 (resp.
1600K for ClueWeb09), the authors report%
~\cite[\nopp Figure~7]{PibiriVenturini2017} a $3\%$ (resp. $1.5\%$) gain in the
index space. In their paper, this percentage takes into account an overhead
that is not present in our paper (to allow block skipping). We thus took this
percentage w.r.t. the value given by their implementation to output the ``CPEF,
estimation'' line in the first part of this table.

In the implementation of \cite{CPEF2017}, we only show the results for the
``docs'', not the ``freqs'' (we had to artificially add freqs to the dataset
\cite{IC2014Dataset} to be able to use \cite{CPEF2017}, we chose to put all
equal to $1$). We note that we modified the Packed+ANS2~\cite{MoffatPetri2018}
implementation to avoid modeling the ``freqs''. Indeed, in the original
implementation, a unique model is used to encode both the ``docs'' and the
``freqs'' part --- the space compression ratio we show is thus better than the
one that you would obtain by using their unmodified implementation.

When removing the ``redundant max'', we instrumented the code to compute the
size of the unnecessary VariableByte encodings, see \figureRef{document_array}. On each dataset, we applied the bisection (b) algorithm%
~\cite{Dhulipala2016, MackenzieMalliaPetriCulpepperSuel2019} or we let
the documents Ids sorted with a lexicographical sorting on the corresponding
URIs (u)~\cite[\nopp Section~5]{SilvestriVenturini2010},
and the resulting integer lists were sorted (s) by increasing length.
\item Finally in Table~\ref{table:interpolative_implems_unfiltered}, we provide this data on unfiltered datasets.
\end{itemize}
\vspace*{-0.3cm}
\begin{table}[!ht]
	\begin{tabularx}{\textwidth}{X*{3}{c}}
		\Xhline{2.5pt}
							   & \mailDatasetOne(b) & TREC(b) & Gov2(b) \\
		\hline
		\Interp                                    & 3.012 & 4.529 & 2.691 \\ 
		Block-\Interp (padding)                    & 3.140 & 4.640 & 2.740 \\ 
		Block-\Interp (no padding)                 & 3.092 & 4.603 & 2.722 \\ 
		Packed+Arithmetic (no padding)             & 2.868 & 4.628 & 2.708 \\ 
		Contextual trits (no padding)              & 2.618 & 4.489 & 2.575 \\ 
		\Interp, clusters (no padding)             & 3.091 & 4.628 & 2.728 \\ 
		\Interp, shrinked clusters (no padding)    & 3.046 & 4.539 & 2.677 \\ 
		EF                                         & 5,514 & 6.091 & 7.871 \\ 
		Block-EF (no padding)                      & 4.708 & 5.309 & 3.762 \\ 
		PEF-approx (no padding)                    & 3.241 & 5.260 & 2.989 \\ 
		PEF-opt (no padding)                       & 3.184 & 4.910 & n/a \\ 
		PEF-approx, clusters (no padding)          & 3.317 & 5.348 & 3.034 \\ 
		PEF-approx, shrinked clusters (no padding) & 3.261 & 5.248 & 2.976 \\ 
		PEF-opt, clusters (no padding)             & 3.243 & 5.009 & n/a \\ 
		PEF-opt, shrinked clusters (no padding)    & 3.191 & 4.916 & n/a \\ 
		\Xhline{2.5pt}
	\end{tabularx}

\caption{Total index size (in bits per integer) for different methods on various datasets.}
\label{table:interpolative_implems}
\end{table}

\vspace*{-0.3cm}
\begin{table}[!ht]
	\begin{tabularx}{\textwidth}{X*{5}{c}}
		\Xhline{2.5pt}
					& Gov2(bf) & Gov2(uf) & ClueWeb09(bf) & ClueWeb09(uf)\\
		\Xhline{2.5pt}
		\Interp					                    & 2.119 & 2.663 & 3.894 & 4.484 \\ 
		Block-\Interp				                & 2.157 & 2.699 & 3.922 & 4.512 \\ 
		PEF					                        & 2.417 & 3.010 & 4.415 & 5.029 \\ 
		Packed+Arithmetic			                & 2.142 & 2.678 & 3.885 & 4.518 \\ 
		Contextual trits			                & 2.054 & 2.558 & 3.749 & 4.307 \\ 
		Clusters-Packed+Arithmetic		            & 2.156 & 2.446 & 3.878 & 4.255 \\ 
		Clusters-\Interp			                & 2.101 & 2.380 & 3.868 & 4.208 \\ 
		Clusters-Contextual trits		            & 2.086 & 2.327 & 3.756 & 4.096 \\ 
		Clusters-PEF				                & 2.400 & 2.714 & 4.392 & 4.748 \\ 
		\hline
		CPEF (``space-based'', estimation)		    & 2.315 & 2.617 & 4.315 & 4.667 \\ 
		\Xhline{2.5pt}
		(Skipping) PEF~\cite{CPEF2017}			    & 2.672 & 3.362 & 4.890 & 5.618 \\ 
		(Skipping) CPEF~\cite{CPEF2017}			    & 2.723 & 3.134 & 5.017 & 5.351 \\ 
		(Skipping) Packed+ANS2~\cite{MoffatPetri2018}	& 2.427 & 2.968 & 4.204 & 4.844 \\ 
		(Skipping) Block-\Interp~\cite{CPEF2017}	& 2.596 & 3.101 & 4.360 & 4.943 \\ 
		\hline
		(Skipping) Block-\Interp (optimized)		& 2.502 & 3.007 & 4.242 & 4.825 \\ 
		\Xhline{2.5pt}
	\end{tabularx}

\caption{Total index size (in bits per integer) on filtered web crawling datasets.}
\label{table:interpolative_implems_filtered}
\end{table}

\begin{table}[!ht]
	\begin{tabularx}{\textwidth}{X*{5}{c}}
		\Xhline{2.5pt}
					& Gov2(b) & Gov2(u) & ClueWeb09(b) & ClueWeb09(u)\\
		\hline
		\Interp					                        & 2.690 & 3.485 & 4.309 & 4.999 \\ 
		Block-\Interp				                    & 2.740 & 3.521 & 4.338 & 5.028 \\ 
		Packed+Arithmetic			                    & 2.708 & 3.462 & 4.274 & 4.982 \\ 
		Contextual trits			                    & 2.575 & 3.311 & 4.119 & 4.763 \\ 
		\hline
		(Skipping) Block-\Interp~\cite{CPEF2017}	    & 3.319 & 4.105 & 4.885 & 5.568 \\ 
		(Skipping) Packed+ANS2~\cite{MoffatPetri2018}	& 2.971 & 3.750 & 4.582 & 5.298 \\ 
		(Skipping) Block-\Interp (optimized)		    & 3.212 & 3.998 & 4.759 & 5.443 \\ 
		\Xhline{2.5pt}
	\end{tabularx}
    \caption{Total index size (in bits per integer) on web crawling datasets.
On each dataset, we applied the bisection (b) algorithm%
~\cite{Dhulipala2016, MackenzieMalliaPetriCulpepperSuel2019} or we let
the documents Ids sorted with a lexicographical sorting on the corresponding
URIs (u)~\cite[\nopp Section~5]{SilvestriVenturini2010}.}
\label{table:interpolative_implems_unfiltered}
\end{table}

\newpage
\section{Implementation details}
\label{appendix:implementation_details}

Our implementation is available at \url{https://gitlab.inria.fr/ybarsami/compress-lists}.
We use the arithmetic encoder from the Reference arithmetic coding library%
~\cite{ArithmeticCoding2011}. We slightly modified the original codebase in
order to more efficiently use adaptive contexts.

All our methods work gap list by gap list. They first read a full gap list from
an index, store it in main memory in an \verb|IntsRef| (it is more or less the
\verb|Java| equivalent of a pointer on an array of \verb|int|s), and then work
on this \verb|IntsRef|. This design choice was made in order to facilitate the
use of the ``SIRENe''~\cite{SIRENe2014} (implementation of the AFOR%
~\cite{DelbruCampinasTummarello2012} method) library that uses this data
structure, and it also facilitates the use of the ``JavaFastPFOR''%
~\cite{JavaFastPFOR2013} library, that uses \verb|int[]| to store the gap
lists.

If one is interested in the running time of the methods implemented in our
repository, it would be more efficient to avoid the use of this intermediate
data structure, that incurs a $2x$ overhead in memory transfers (first, read
from file and store in the array, then read from this array and do the
computations).

The implementation of the bit lists, trit lists and quatrit lists all use the
\verb|java.util.BitSet| data structure internally, in order to be memory
efficient. This is not required as those lists may be constructed on the fly
instead of being stored in main memory --- there is another $2x$ overhead in
memory transfers behind this choice.

All the methods are implemented inside the \verb|gaparrayencoding| folder.

\section{Variants of our method}\label{appendix:C}
\subsection{More subtle contexts}

At the beginning of our work is the empirical establishment that contexts are
very useful for entropy coding. We then searched a balance between the
precision of the contexts used and the size of the probabilities needed to
store those contexts.

The most efficient compromise we found was to divide a context of size $\gamma$
into the $k$ last trits, and the $w$ trits before. On the $k$ last trits, what
is looked at is whether each trit is a $2$ or not, and on the $w$ trits before,
what is looked at is the number of $2$ in total.

We also tested to have a larger number of sub-boxes inside the $\gamma$ last
trits, and it did not turn out to be useful. For instance, we would divide the
$\gamma$ last trits in the $k$ last trits, where we would, as before, take the
full context (just merging $0$s and $1$s), then look at the $w_0$ previous
trits, only counting the number of $2$ in it, then the $w_1$ trits before, also
only counting the number of $2$ in it, etc.

In a way, our experiments are a specialized version of the more general
GRASP algorithm pictured in \cite{MrakMarpeWiegand2003}, which also tries to
merge contexts in order to get information from a greater context without
having to pay for the full price. It would be interesting to see if this more
general idea could lead to better compression rates, and to analyze the kind
of contexts that would be merged together.

\subsection{Batching}
\label{section:batching}

We can refine our contextual methods with batching. We experimented and found
that sometimes, a logarithmic batching (with respect to the gap list lengths)
gives slightly better results (on half of the datasets), but the gains were
marginal. We used $7$~batches and use, for each batch, the best method between
\Interp, TC and TCA. Our batches have the following delimiters:
$$\{0,\enspace 0.0002,\enspace 0.001,\enspace 0.01,\enspace 0.03,\enspace 0.1,\enspace 0.2,\enspace 1\}$$
in terms of gap array density (gap array lengths divided by $\nbD$).

When using the Bookstein method, on the other hand, batching was more than
needed: for instance on the Bible dataset, using the Bookstein method with
batches gives a $25\%$ improvement on the index size compared to the
Bookstein method without batches; on the Enron dataset, the gain was $31\%$.
We can see the Bookstein method as a method with only $2$~bits of context,
while our method use larger contexts. In a way, the high number of contexts we
use is more or less giving the information about the density, which explains
why batching is not fundamental for our method.

\subsection{Quatritlist}

In all previous papers, and in previous sections, the index was always created
with an explicit storage of the gap array lengths. For most methods, this
knowledge is mandatory. But in our case, we can avoid storing the gap array
lengths. When using arithmetic coding, in order to decode we need a way of
knowing when to stop decoding. Instead of using the gap array lengths, we can
thus encode a special ``end'' character, at the end of each gap array. If we
note this special character ``3'', and we are now dealing with quatrits
(integers in $\{0, 1, 2, 3\}$).

We tested all our methods with quatrits, and the results are equivalent to the
results with the explicit storage of the gap array lengths. However, if one is
dealing with a dataset for which this storage of lengths is higher, this method
surely helps.

\subsection{Context on the bit vector}

In \sectionRef{related}, we noted that previous works on inverted index
compression also used contextual encodings, \eg, contextual arithmetic encoding
on bit vectors~\cite{BooksteinKleinRaita1997}. We thus also tested contextual
encodings on the bit vector.
Our experiments showed that these methods are outperformed by our trits ---
even though contextual encodings on the bit vector are also outperforming
both \Interp and Packed+Arithmetic methods.
It would be interesting to see if a more careful setting of parameters would
lead to better results.

\section{Influence of parameters on the TREC dataset}\label{appendix:D}

The results presented in this appendix are very similar to those presented in
\sectionRef{interpretation} about the ENRON dataset, therefore the
interpretation of those tables stays the same. In \sectionRef{interpretation},
it was clear that the crucial optimization was to merge the symbols ``$0$'' and
``$1$'' (the context can be seen as ``$T$'' for the ``$2$'' symbol and ``$N$''
for both symbols ``$0$'' and ``$1$''), so we this time present only two
scenarios:

\begin{itemize}
	\item using static contextual encoding, merging the symbols ``$0$'' and
	``$1$'' and using $8$~bits for storing probabilities;
	\item using adaptive contextual encoding.
\end{itemize}

In each of those tables, the three bold values represent the three smallest
values. While comparing these tables with the ones from the ENRON dataset, we
can see that the optimal values are the same for $k$, but bigger for $w$. In a
way, the datasets do not differ much (the TREC dataset contains approximately
twice the number of documents, of words, and of integers to compress), so it
is good to see that similar values for $k$ and $w$ give very good compression
results. The size to store the list lengths is not accounted for in those
results. It is of course constant, equal to $0,380$~MB for the TREC dataset.

\begin{landscape}
\begin{table}[!ht]
\setlength\tabcolsep{1pt}
\renewcommand{\arraystretch}{1.5}

\begin{scriptsize}
\noindent\begin{tabularx}{\linewidth}{|l|*{16}{>{\centering \arraybackslash}X|}} \hline
 & $w = 0$ & $w = 4$ & $w = 8$ & $w = 12$ & $w = 16$ & $w = 20$ & $w = 24$ & $w = 28$ & $w = 32$ & $w = 36$ & $w = 40$ & $w = 44$ & $w = 48$ & $w = 52$ & $w = 56$ & $w = 60$ \\\hline
$k = 0$ & \cellcolor{MyColor239!50!white}$79.02_{0.00}^{79.02}$ & \cellcolor{MyColor221!50!white}$74.33_{0.00}^{74.33}$ & \cellcolor{MyColor203!50!white}$72.87_{0.00}^{72.87}$ & \cellcolor{MyColor191!50!white}$72.14_{0.00}^{72.13}$ & \cellcolor{MyColor182!50!white}$71.69_{0.00}^{71.68}$ & \cellcolor{MyColor184!50!white}$71.75_{0.00}^{71.75}$ & \cellcolor{MyColor181!50!white}$71.60_{0.00}^{71.60}$ & \cellcolor{MyColor178!50!white}$71.49_{0.00}^{71.49}$ & \cellcolor{MyColor177!50!white}$71.42_{0.00}^{71.42}$ & \cellcolor{MyColor176!50!white}$71.39_{0.00}^{71.39}$ & \cellcolor{MyColor175!50!white}$71.35_{0.00}^{71.35}$ & \cellcolor{MyColor175!50!white}$71.32_{0.00}^{71.32}$ & \cellcolor{MyColor174!50!white}$71.30_{0.00}^{71.30}$ & \cellcolor{MyColor174!50!white}$71.29_{0.00}^{71.29}$ & \cellcolor{MyColor174!50!white}$71.28_{0.00}^{71.28}$ & \cellcolor{MyColor174!50!white}$71.27_{0.00}^{71.27}$ \\\hline
$k = 1$ & \cellcolor{MyColor239!50!white}$78.52_{0.00}^{78.52}$ & \cellcolor{MyColor204!50!white}$72.91_{0.00}^{72.91}$ & \cellcolor{MyColor177!50!white}$71.43_{0.00}^{71.43}$ & \cellcolor{MyColor160!50!white}$70.74_{0.00}^{70.74}$ & \cellcolor{MyColor149!50!white}$70.34_{0.00}^{70.34}$ & \cellcolor{MyColor153!50!white}$70.46_{0.00}^{70.46}$ & \cellcolor{MyColor149!50!white}$70.34_{0.00}^{70.33}$ & \cellcolor{MyColor146!50!white}$70.26_{0.00}^{70.25}$ & \cellcolor{MyColor145!50!white}$70.21_{0.00}^{70.21}$ & \cellcolor{MyColor145!50!white}$70.21_{0.00}^{70.20}$ & \cellcolor{MyColor144!50!white}$70.18_{0.00}^{70.17}$ & \cellcolor{MyColor144!50!white}$70.16_{0.00}^{70.16}$ & \cellcolor{MyColor143!50!white}$70.16_{0.00}^{70.15}$ & \cellcolor{MyColor143!50!white}$70.16_{0.00}^{70.16}$ & \cellcolor{MyColor143!50!white}$70.15_{0.00}^{70.15}$ & \cellcolor{MyColor143!50!white}$70.16_{0.00}^{70.16}$ \\\hline
$k = 2$ & \cellcolor{MyColor234!50!white}$76.16_{0.00}^{76.16}$ & \cellcolor{MyColor184!50!white}$71.76_{0.00}^{71.76}$ & \cellcolor{MyColor148!50!white}$70.32_{0.00}^{70.32}$ & \cellcolor{MyColor126!50!white}$69.65_{0.00}^{69.65}$ & \cellcolor{MyColor113!50!white}$69.30_{0.00}^{69.30}$ & \cellcolor{MyColor118!50!white}$69.42_{0.00}^{69.42}$ & \cellcolor{MyColor113!50!white}$69.31_{0.00}^{69.31}$ & \cellcolor{MyColor110!50!white}$69.24_{0.00}^{69.24}$ & \cellcolor{MyColor109!50!white}$69.21_{0.00}^{69.21}$ & \cellcolor{MyColor109!50!white}$69.20_{0.00}^{69.20}$ & \cellcolor{MyColor108!50!white}$69.18_{0.00}^{69.18}$ & \cellcolor{MyColor108!50!white}$69.18_{0.00}^{69.17}$ & \cellcolor{MyColor108!50!white}$69.17_{0.00}^{69.17}$ & \cellcolor{MyColor108!50!white}$69.18_{0.00}^{69.18}$ & \cellcolor{MyColor108!50!white}$69.18_{0.00}^{69.18}$ & \cellcolor{MyColor108!50!white}$69.18_{0.00}^{69.18}$ \\\hline
$k = 3$ & \cellcolor{MyColor223!50!white}$74.58_{0.00}^{74.58}$ & \cellcolor{MyColor165!50!white}$70.92_{0.00}^{70.92}$ & \cellcolor{MyColor123!50!white}$69.56_{0.00}^{69.56}$ & \cellcolor{MyColor97!50!white}$68.94_{0.00}^{68.93}$ & \cellcolor{MyColor96!50!white}$68.91_{0.00}^{68.91}$ & \cellcolor{MyColor89!50!white}$68.75_{0.00}^{68.75}$ & \cellcolor{MyColor84!50!white}$68.65_{0.00}^{68.64}$ & \cellcolor{MyColor81!50!white}$68.58_{0.00}^{68.58}$ & \cellcolor{MyColor80!50!white}$68.56_{0.00}^{68.56}$ & \cellcolor{MyColor79!50!white}$68.55_{0.00}^{68.55}$ & \cellcolor{MyColor78!50!white}$68.53_{0.00}^{68.53}$ & \cellcolor{MyColor78!50!white}$68.53_{0.00}^{68.52}$ & \cellcolor{MyColor78!50!white}$68.53_{0.00}^{68.53}$ & \cellcolor{MyColor78!50!white}$68.53_{0.00}^{68.53}$ & \cellcolor{MyColor78!50!white}$68.53_{0.00}^{68.53}$ & \cellcolor{MyColor79!50!white}$68.54_{0.00}^{68.54}$ \\\hline
$k = 4$ & \cellcolor{MyColor210!50!white}$73.35_{0.00}^{73.35}$ & \cellcolor{MyColor148!50!white}$70.29_{0.00}^{70.29}$ & \cellcolor{MyColor102!50!white}$69.05_{0.00}^{69.05}$ & \cellcolor{MyColor75!50!white}$68.47_{0.00}^{68.46}$ & \cellcolor{MyColor76!50!white}$68.49_{0.00}^{68.49}$ & \cellcolor{MyColor68!50!white}$68.34_{0.00}^{68.33}$ & \cellcolor{MyColor63!50!white}$68.23_{0.00}^{68.23}$ & \cellcolor{MyColor59!50!white}$68.17_{0.00}^{68.16}$ & \cellcolor{MyColor58!50!white}$68.16_{0.00}^{68.15}$ & \cellcolor{MyColor57!50!white}$68.14_{0.00}^{68.13}$ & \cellcolor{MyColor56!50!white}$68.12_{0.00}^{68.11}$ & \cellcolor{MyColor56!50!white}$68.11_{0.00}^{68.11}$ & \cellcolor{MyColor56!50!white}$68.11_{0.00}^{68.11}$ & \cellcolor{MyColor56!50!white}$68.11_{0.00}^{68.11}$ & \cellcolor{MyColor56!50!white}$68.12_{0.00}^{68.11}$ & \cellcolor{MyColor56!50!white}$68.12_{0.00}^{68.12}$ \\\hline
$k = 5$ & \cellcolor{MyColor195!50!white}$72.35_{0.00}^{72.35}$ & \cellcolor{MyColor131!50!white}$69.79_{0.00}^{69.78}$ & \cellcolor{MyColor86!50!white}$68.68_{0.00}^{68.68}$ & \cellcolor{MyColor58!50!white}$68.15_{0.00}^{68.15}$ & \cellcolor{MyColor62!50!white}$68.22_{0.00}^{68.21}$ & \cellcolor{MyColor53!50!white}$68.07_{0.00}^{68.06}$ & \cellcolor{MyColor47!50!white}$67.96_{0.00}^{67.96}$ & \cellcolor{MyColor44!50!white}$67.90_{0.00}^{67.90}$ & \cellcolor{MyColor43!50!white}$67.89_{0.00}^{67.89}$ & \cellcolor{MyColor42!50!white}$67.87_{0.00}^{67.86}$ & \cellcolor{MyColor40!50!white}$67.85_{0.00}^{67.84}$ & \cellcolor{MyColor40!50!white}$67.84_{0.00}^{67.83}$ & \cellcolor{MyColor40!50!white}$67.84_{0.00}^{67.84}$ & \cellcolor{MyColor40!50!white}$67.84_{0.00}^{67.83}$ & \cellcolor{MyColor40!50!white}$67.84_{0.01}^{67.83}$ & \cellcolor{MyColor40!50!white}$67.84_{0.01}^{67.84}$ \\\hline
$k = 6$ & \cellcolor{MyColor178!50!white}$71.49_{0.00}^{71.49}$ & \cellcolor{MyColor115!50!white}$69.37_{0.00}^{69.37}$ & \cellcolor{MyColor71!50!white}$68.39_{0.00}^{68.39}$ & \cellcolor{MyColor46!50!white}$67.93_{0.00}^{67.93}$ & \cellcolor{MyColor50!50!white}$68.00_{0.00}^{68.00}$ & \cellcolor{MyColor41!50!white}$67.86_{0.00}^{67.85}$ & \cellcolor{MyColor35!50!white}$67.75_{0.00}^{67.75}$ & \cellcolor{MyColor31!50!white}$67.70_{0.01}^{67.69}$ & \cellcolor{MyColor31!50!white}$67.69_{0.01}^{67.68}$ & \cellcolor{MyColor29!50!white}$67.66_{0.01}^{67.65}$ & \cellcolor{MyColor27!50!white}$67.64_{0.01}^{67.63}$ & \cellcolor{MyColor27!50!white}$67.63_{0.01}^{67.62}$ & \cellcolor{MyColor27!50!white}$67.63_{0.01}^{67.62}$ & \cellcolor{MyColor26!50!white}$67.62_{0.01}^{67.62}$ & \cellcolor{MyColor26!50!white}$67.62_{0.01}^{67.61}$ & \cellcolor{MyColor27!50!white}$67.63_{0.01}^{67.62}$ \\\hline
$k = 7$ & \cellcolor{MyColor161!50!white}$70.76_{0.00}^{70.76}$ & \cellcolor{MyColor100!50!white}$69.00_{0.00}^{69.00}$ & \cellcolor{MyColor58!50!white}$68.14_{0.00}^{68.14}$ & \cellcolor{MyColor51!50!white}$68.03_{0.00}^{68.03}$ & \cellcolor{MyColor40!50!white}$67.84_{0.01}^{67.83}$ & \cellcolor{MyColor31!50!white}$67.70_{0.01}^{67.69}$ & \cellcolor{MyColor24!50!white}$67.59_{0.01}^{67.58}$ & \cellcolor{MyColor21!50!white}$67.54_{0.01}^{67.53}$ & \cellcolor{MyColor20!50!white}$67.53_{0.01}^{67.52}$ & \cellcolor{MyColor18!50!white}$67.49_{0.01}^{67.48}$ & \cellcolor{MyColor17!50!white}$67.47_{0.01}^{67.46}$ & \cellcolor{MyColor16!50!white}$67.47_{0.01}^{67.45}$ & \cellcolor{MyColor16!50!white}$67.46_{0.01}^{67.45}$ & \cellcolor{MyColor15!50!white}$67.46_{0.02}^{67.44}$ & \cellcolor{MyColor15!50!white}$67.46_{0.02}^{67.44}$ & \cellcolor{MyColor15!50!white}$67.46_{0.02}^{67.44}$ \\\hline
$k = 8$ & \cellcolor{MyColor143!50!white}$70.14_{0.00}^{70.14}$ & \cellcolor{MyColor85!50!white}$68.68_{0.00}^{68.67}$ & \cellcolor{MyColor45!50!white}$67.92_{0.01}^{67.92}$ & \cellcolor{MyColor42!50!white}$67.88_{0.01}^{67.87}$ & \cellcolor{MyColor32!50!white}$67.70_{0.01}^{67.69}$ & \cellcolor{MyColor23!50!white}$67.57_{0.01}^{67.55}$ & \cellcolor{MyColor16!50!white}$67.46_{0.01}^{67.45}$ & \cellcolor{MyColor14!50!white}$67.43_{0.02}^{67.42}$ & \cellcolor{MyColor12!50!white}$67.40_{0.02}^{67.39}$ & \cellcolor{MyColor10!50!white}$67.37_{0.02}^{67.35}$ & \cellcolor{MyColor8!50!white}$67.35_{0.02}^{67.33}$ & \cellcolor{MyColor8!50!white}$67.35_{0.02}^{67.32}$ & \cellcolor{MyColor7!50!white}$67.34_{0.03}^{67.32}$ & \cellcolor{MyColor7!50!white}$67.34_{0.03}^{67.31}$ & \cellcolor{MyColor7!50!white}$67.33_{0.03}^{67.30}$ & \cellcolor{MyColor7!50!white}$67.34_{0.03}^{67.30}$ \\\hline
$k = 9$ & \cellcolor{MyColor125!50!white}$69.63_{0.00}^{69.62}$ & \cellcolor{MyColor71!50!white}$68.39_{0.01}^{68.38}$ & \cellcolor{MyColor34!50!white}$67.74_{0.01}^{67.73}$ & \cellcolor{MyColor35!50!white}$67.75_{0.01}^{67.74}$ & \cellcolor{MyColor25!50!white}$67.60_{0.02}^{67.58}$ & \cellcolor{MyColor16!50!white}$67.47_{0.02}^{67.44}$ & \cellcolor{MyColor10!50!white}$67.37_{0.03}^{67.35}$ & \cellcolor{MyColor8!50!white}$67.35_{0.03}^{67.32}$ & \cellcolor{MyColor6!50!white}$67.32_{0.04}^{67.28}$ & \cellcolor{MyColor4!50!white}$67.29_{0.04}^{67.25}$ & \cellcolor{MyColor2!50!white}$67.27_{0.04}^{67.23}$ & \cellcolor{MyColor2!50!white}$67.27_{0.05}^{67.22}$ & \cellcolor{MyColor2!50!white}$67.26_{0.05}^{67.21}$ & \cellcolor{MyColor1!50!white}$67.26_{0.06}^{67.20}$ & \cellcolor{MyColor1!50!white}$67.26_{0.06}^{67.20}$ & \cellcolor{MyColor2!50!white}$67.26_{0.06}^{67.20}$ \\\hline
$k = 10$ & \cellcolor{MyColor109!50!white}$69.20_{0.00}^{69.20}$ & \cellcolor{MyColor58!50!white}$68.15_{0.01}^{68.14}$ & \cellcolor{MyColor26!50!white}$67.61_{0.02}^{67.59}$ & \cellcolor{MyColor28!50!white}$67.65_{0.03}^{67.62}$ & \cellcolor{MyColor20!50!white}$67.52_{0.04}^{67.48}$ & \cellcolor{MyColor11!50!white}$67.40_{0.04}^{67.35}$ & \cellcolor{MyColor6!50!white}$67.31_{0.05}^{67.26}$ & \cellcolor{MyColor4!50!white}$67.30_{0.06}^{67.24}$ & \cellcolor{MyColor2!50!white}$67.27_{0.07}^{67.20}$ & \cellcolor{MyColor0!50!white}$67.24_{0.08}^{67.17}$ & \cellcolor{MyColor0!50!white}$67.23_{0.09}^{67.14}$ & \cellcolor{MyColor0!50!white}$67.23_{0.09}^{67.14}$ & \cellcolor{MyColor0!50!white}$\mathbf{67.23_{0.10}^{67.13}}$ & \cellcolor{MyColor0!50!white}$\mathbf{67.23_{0.11}^{67.12}}$ & \cellcolor{MyColor0!50!white}$\mathbf{67.23_{0.12}^{67.11}}$ & \cellcolor{MyColor0!50!white}$67.24_{0.13}^{67.11}$ \\\hline
$k = 11$ & \cellcolor{MyColor93!50!white}$68.84_{0.01}^{68.84}$ & \cellcolor{MyColor46!50!white}$67.94_{0.02}^{67.91}$ & \cellcolor{MyColor37!50!white}$67.79_{0.04}^{67.75}$ & \cellcolor{MyColor23!50!white}$67.58_{0.05}^{67.52}$ & \cellcolor{MyColor16!50!white}$67.46_{0.07}^{67.39}$ & \cellcolor{MyColor8!50!white}$67.36_{0.09}^{67.27}$ & \cellcolor{MyColor4!50!white}$67.29_{0.10}^{67.19}$ & \cellcolor{MyColor3!50!white}$67.28_{0.12}^{67.16}$ & \cellcolor{MyColor2!50!white}$67.26_{0.14}^{67.12}$ & \cellcolor{MyColor1!50!white}$67.24_{0.15}^{67.09}$ & \cellcolor{MyColor0!50!white}$67.24_{0.17}^{67.07}$ & \cellcolor{MyColor1!50!white}$67.25_{0.19}^{67.06}$ & \cellcolor{MyColor1!50!white}$67.25_{0.20}^{67.05}$ & \cellcolor{MyColor2!50!white}$67.26_{0.22}^{67.04}$ & \cellcolor{MyColor2!50!white}$67.27_{0.23}^{67.03}$ & \cellcolor{MyColor3!50!white}$67.28_{0.25}^{67.03}$ \\\hline
$k = 12$ & \cellcolor{MyColor78!50!white}$68.53_{0.01}^{68.52}$ & \cellcolor{MyColor34!50!white}$67.75_{0.04}^{67.71}$ & \cellcolor{MyColor32!50!white}$67.71_{0.08}^{67.63}$ & \cellcolor{MyColor21!50!white}$67.54_{0.11}^{67.43}$ & \cellcolor{MyColor15!50!white}$67.45_{0.14}^{67.31}$ & \cellcolor{MyColor9!50!white}$67.36_{0.17}^{67.19}$ & \cellcolor{MyColor7!50!white}$67.33_{0.21}^{67.13}$ & \cellcolor{MyColor6!50!white}$67.33_{0.24}^{67.09}$ & \cellcolor{MyColor6!50!white}$67.32_{0.27}^{67.05}$ & \cellcolor{MyColor6!50!white}$67.32_{0.30}^{67.02}$ & \cellcolor{MyColor7!50!white}$67.34_{0.34}^{67.00}$ & \cellcolor{MyColor9!50!white}$67.36_{0.37}^{66.99}$ & \cellcolor{MyColor10!50!white}$67.38_{0.40}^{66.97}$ & \cellcolor{MyColor11!50!white}$67.40_{0.44}^{66.96}$ & \cellcolor{MyColor13!50!white}$67.42_{0.47}^{66.96}$ & \cellcolor{MyColor15!50!white}$67.45_{0.50}^{66.95}$ \\\hline
$k = 13$ & \cellcolor{MyColor64!50!white}$68.26_{0.02}^{68.24}$ & \cellcolor{MyColor25!50!white}$67.60_{0.08}^{67.52}$ & \cellcolor{MyColor30!50!white}$67.67_{0.15}^{67.52}$ & \cellcolor{MyColor22!50!white}$67.56_{0.21}^{67.34}$ & \cellcolor{MyColor18!50!white}$67.50_{0.28}^{67.22}$ & \cellcolor{MyColor15!50!white}$67.45_{0.35}^{67.11}$ & \cellcolor{MyColor16!50!white}$67.47_{0.41}^{67.06}$ & \cellcolor{MyColor17!50!white}$67.48_{0.48}^{67.01}$ & \cellcolor{MyColor19!50!white}$67.51_{0.54}^{66.97}$ & \cellcolor{MyColor21!50!white}$67.55_{0.61}^{66.94}$ & \cellcolor{MyColor25!50!white}$67.59_{0.67}^{66.92}$ & \cellcolor{MyColor28!50!white}$67.64_{0.74}^{66.90}$ & \cellcolor{MyColor31!50!white}$67.69_{0.80}^{66.89}$ & \cellcolor{MyColor34!50!white}$67.74_{0.87}^{66.87}$ & \cellcolor{MyColor38!50!white}$67.80_{0.94}^{66.86}$ & \cellcolor{MyColor41!50!white}$67.85_{1.00}^{66.85}$ \\\hline
$k = 14$ & \cellcolor{MyColor52!50!white}$68.05_{0.03}^{68.01}$ & \cellcolor{MyColor21!50!white}$67.54_{0.17}^{67.38}$ & \cellcolor{MyColor32!50!white}$67.71_{0.30}^{67.42}$ & \cellcolor{MyColor30!50!white}$67.68_{0.43}^{67.25}$ & \cellcolor{MyColor30!50!white}$67.68_{0.56}^{67.12}$ & \cellcolor{MyColor32!50!white}$67.70_{0.69}^{67.02}$ & \cellcolor{MyColor37!50!white}$67.79_{0.82}^{66.97}$ & \cellcolor{MyColor42!50!white}$67.87_{0.95}^{66.91}$ & \cellcolor{MyColor47!50!white}$67.96_{1.08}^{66.87}$ & \cellcolor{MyColor53!50!white}$68.05_{1.21}^{66.84}$ & \cellcolor{MyColor59!50!white}$68.16_{1.35}^{66.82}$ & \cellcolor{MyColor65!50!white}$68.27_{1.48}^{66.79}$ & \cellcolor{MyColor70!50!white}$68.38_{1.61}^{66.77}$ & \cellcolor{MyColor76!50!white}$68.49_{1.74}^{66.75}$ & \cellcolor{MyColor82!50!white}$68.61_{1.87}^{66.74}$ & \cellcolor{MyColor88!50!white}$68.73_{2.00}^{66.73}$ \\\hline
$k = 15$ & \cellcolor{MyColor41!50!white}$67.86_{0.07}^{67.79}$ & \cellcolor{MyColor41!50!white}$67.85_{0.33}^{67.52}$ & \cellcolor{MyColor44!50!white}$67.90_{0.59}^{67.31}$ & \cellcolor{MyColor49!50!white}$67.99_{0.85}^{67.14}$ & \cellcolor{MyColor57!50!white}$68.12_{1.12}^{67.01}$ & \cellcolor{MyColor65!50!white}$68.28_{1.38}^{66.90}$ & \cellcolor{MyColor76!50!white}$68.49_{1.64}^{66.85}$ & \cellcolor{MyColor86!50!white}$68.69_{1.90}^{66.79}$ & \cellcolor{MyColor96!50!white}$68.90_{2.16}^{66.74}$ & \cellcolor{MyColor106!50!white}$69.13_{2.43}^{66.70}$ & \cellcolor{MyColor115!50!white}$69.36_{2.69}^{66.67}$ & \cellcolor{MyColor124!50!white}$69.59_{2.95}^{66.64}$ & \cellcolor{MyColor132!50!white}$69.82_{3.21}^{66.61}$ & \cellcolor{MyColor140!50!white}$70.06_{3.47}^{66.58}$ & \cellcolor{MyColor148!50!white}$70.29_{3.74}^{66.56}$ & \cellcolor{MyColor155!50!white}$70.53_{4.00}^{66.53}$ \\\hline
$k = 16$ & \cellcolor{MyColor33!50!white}$67.73_{0.13}^{67.59}$ & \cellcolor{MyColor52!50!white}$68.04_{0.66}^{67.39}$ & \cellcolor{MyColor69!50!white}$68.36_{1.18}^{67.17}$ & \cellcolor{MyColor87!50!white}$68.71_{1.71}^{67.00}$ & \cellcolor{MyColor104!50!white}$69.08_{2.23}^{66.85}$ & \cellcolor{MyColor121!50!white}$69.51_{2.75}^{66.76}$ & \cellcolor{MyColor137!50!white}$69.95_{3.28}^{66.68}$ & \cellcolor{MyColor151!50!white}$70.40_{3.80}^{66.60}$ & \cellcolor{MyColor164!50!white}$70.86_{4.33}^{66.53}$ & \cellcolor{MyColor175!50!white}$71.34_{4.85}^{66.48}$ & \cellcolor{MyColor185!50!white}$71.81_{5.38}^{66.44}$ & \cellcolor{MyColor194!50!white}$72.29_{5.90}^{66.39}$ & \cellcolor{MyColor202!50!white}$72.77_{6.42}^{66.34}$ & \cellcolor{MyColor208!50!white}$73.25_{6.95}^{66.31}$ & \cellcolor{MyColor214!50!white}$73.74_{7.47}^{66.27}$ & \cellcolor{MyColor219!50!white}$74.23_{8.00}^{66.23}$ \\\hline
$k = 17$ & \cellcolor{MyColor30!50!white}$67.67_{0.26}^{67.41}$ & \cellcolor{MyColor78!50!white}$68.54_{1.31}^{67.22}$ & \cellcolor{MyColor115!50!white}$69.36_{2.36}^{67.00}$ & \cellcolor{MyColor145!50!white}$70.21_{3.41}^{66.80}$ & \cellcolor{MyColor169!50!white}$71.09_{4.46}^{66.63}$ & \cellcolor{MyColor189!50!white}$72.02_{5.51}^{66.52}$ & \cellcolor{MyColor204!50!white}$72.96_{6.56}^{66.40}$ & \cellcolor{MyColor216!50!white}$73.90_{7.60}^{66.30}$ & \cellcolor{MyColor225!50!white}$74.86_{8.65}^{66.21}$ & \cellcolor{MyColor232!50!white}$75.84_{9.70}^{66.14}$ & \cellcolor{MyColor238!50!white}$76.81_{10.75}^{66.06}$ & \cellcolor{MyColor239!50!white}$77.79_{11.80}^{65.99}$ & \cellcolor{MyColor239!50!white}$78.77_{12.85}^{65.92}$ & \cellcolor{MyColor239!50!white}$79.76_{13.90}^{65.86}$ & \cellcolor{MyColor239!50!white}$80.74_{14.94}^{65.80}$ & \cellcolor{MyColor239!50!white}$81.73_{15.99}^{65.74}$ \\\hline
\end{tabularx}
\end{scriptsize}
\caption{Size of compressed lists for full TREC dataset, without distinction between ``$0$'' and ``$1$'' symbols and $8$~bit probabilities. For example, $68.05_{0.03}^{68.01}$ means that the compressed lists use $68.05$~MB, divided into $0.03$~MB for storing the probabilities $P(i \mid C)$ and $68.01$~MB for the rest. All numbers are given $\pm 0.01$~MB.}
\label{table:10}
\end{table}

\bigskip

\begin{table}[!ht]

\setlength\tabcolsep{1pt}
\renewcommand{\arraystretch}{1.5}

\begin{small}
\noindent\begin{tabularx}{\linewidth}{|l|*{16}{>{\centering \arraybackslash}X|}} \hline
 & $w = 0$ & $w = 4$ & $w = 8$ & $w = 12$ & $w = 16$ & $w = 20$ & $w = 24$ & $w = 28$ & $w = 32$ & $w = 36$ & $w = 40$ & $w = 44$ & $w = 48$ & $w = 52$ & $w = 56$ & $w = 60$ \\\hline
$k = 0$ & \cellcolor{MyColor179!50!white}$71.50$ & \cellcolor{MyColor119!50!white}$69.45$ & \cellcolor{MyColor139!50!white}$70.02$ & \cellcolor{MyColor148!50!white}$70.29$ & \cellcolor{MyColor151!50!white}$70.42$ & \cellcolor{MyColor164!50!white}$70.89$ & \cellcolor{MyColor166!50!white}$70.95$ & \cellcolor{MyColor167!50!white}$71.01$ & \cellcolor{MyColor169!50!white}$71.07$ & \cellcolor{MyColor170!50!white}$71.13$ & \cellcolor{MyColor171!50!white}$71.16$ & \cellcolor{MyColor172!50!white}$71.19$ & \cellcolor{MyColor172!50!white}$71.22$ & \cellcolor{MyColor173!50!white}$71.24$ & \cellcolor{MyColor173!50!white}$71.26$ & \cellcolor{MyColor174!50!white}$71.28$ \\\hline
$k = 1$ & \cellcolor{MyColor155!50!white}$70.54$ & \cellcolor{MyColor100!50!white}$69.00$ & \cellcolor{MyColor109!50!white}$69.22$ & \cellcolor{MyColor112!50!white}$69.28$ & \cellcolor{MyColor114!50!white}$69.33$ & \cellcolor{MyColor129!50!white}$69.74$ & \cellcolor{MyColor130!50!white}$69.77$ & \cellcolor{MyColor132!50!white}$69.81$ & \cellcolor{MyColor134!50!white}$69.86$ & \cellcolor{MyColor136!50!white}$69.92$ & \cellcolor{MyColor136!50!white}$69.94$ & \cellcolor{MyColor137!50!white}$69.96$ & \cellcolor{MyColor138!50!white}$69.99$ & \cellcolor{MyColor139!50!white}$70.02$ & \cellcolor{MyColor140!50!white}$70.04$ & \cellcolor{MyColor140!50!white}$70.06$ \\\hline
$k = 2$ & \cellcolor{MyColor134!50!white}$69.86$ & \cellcolor{MyColor83!50!white}$68.63$ & \cellcolor{MyColor81!50!white}$68.59$ & \cellcolor{MyColor78!50!white}$68.52$ & \cellcolor{MyColor78!50!white}$68.52$ & \cellcolor{MyColor92!50!white}$68.83$ & \cellcolor{MyColor92!50!white}$68.83$ & \cellcolor{MyColor93!50!white}$68.85$ & \cellcolor{MyColor95!50!white}$68.89$ & \cellcolor{MyColor97!50!white}$68.92$ & \cellcolor{MyColor97!50!white}$68.94$ & \cellcolor{MyColor98!50!white}$68.96$ & \cellcolor{MyColor99!50!white}$68.98$ & \cellcolor{MyColor100!50!white}$69.00$ & \cellcolor{MyColor101!50!white}$69.02$ & \cellcolor{MyColor102!50!white}$69.04$ \\\hline
$k = 3$ & \cellcolor{MyColor121!50!white}$69.51$ & \cellcolor{MyColor71!50!white}$68.39$ & \cellcolor{MyColor60!50!white}$68.18$ & \cellcolor{MyColor51!50!white}$68.03$ & \cellcolor{MyColor64!50!white}$68.26$ & \cellcolor{MyColor62!50!white}$68.21$ & \cellcolor{MyColor60!50!white}$68.19$ & \cellcolor{MyColor61!50!white}$68.20$ & \cellcolor{MyColor62!50!white}$68.22$ & \cellcolor{MyColor63!50!white}$68.24$ & \cellcolor{MyColor64!50!white}$68.25$ & \cellcolor{MyColor64!50!white}$68.27$ & \cellcolor{MyColor65!50!white}$68.28$ & \cellcolor{MyColor66!50!white}$68.30$ & \cellcolor{MyColor67!50!white}$68.31$ & \cellcolor{MyColor68!50!white}$68.33$ \\\hline
$k = 4$ & \cellcolor{MyColor114!50!white}$69.33$ & \cellcolor{MyColor62!50!white}$68.21$ & \cellcolor{MyColor42!50!white}$67.87$ & \cellcolor{MyColor29!50!white}$67.66$ & \cellcolor{MyColor40!50!white}$67.84$ & \cellcolor{MyColor36!50!white}$67.78$ & \cellcolor{MyColor34!50!white}$67.75$ & \cellcolor{MyColor34!50!white}$67.74$ & \cellcolor{MyColor36!50!white}$67.77$ & \cellcolor{MyColor35!50!white}$67.77$ & \cellcolor{MyColor36!50!white}$67.77$ & \cellcolor{MyColor36!50!white}$67.78$ & \cellcolor{MyColor37!50!white}$67.79$ & \cellcolor{MyColor38!50!white}$67.80$ & \cellcolor{MyColor38!50!white}$67.81$ & \cellcolor{MyColor39!50!white}$67.82$ \\\hline
$k = 5$ & \cellcolor{MyColor106!50!white}$69.13$ & \cellcolor{MyColor51!50!white}$68.03$ & \cellcolor{MyColor25!50!white}$67.61$ & \cellcolor{MyColor10!50!white}$67.38$ & \cellcolor{MyColor20!50!white}$67.53$ & \cellcolor{MyColor15!50!white}$67.45$ & \cellcolor{MyColor13!50!white}$67.42$ & \cellcolor{MyColor12!50!white}$67.41$ & \cellcolor{MyColor13!50!white}$67.42$ & \cellcolor{MyColor12!50!white}$67.41$ & \cellcolor{MyColor12!50!white}$67.41$ & \cellcolor{MyColor13!50!white}$67.42$ & \cellcolor{MyColor13!50!white}$67.43$ & \cellcolor{MyColor14!50!white}$67.43$ & \cellcolor{MyColor14!50!white}$67.44$ & \cellcolor{MyColor15!50!white}$67.45$ \\\hline
$k = 6$ & \cellcolor{MyColor93!50!white}$68.83$ & \cellcolor{MyColor39!50!white}$67.82$ & \cellcolor{MyColor11!50!white}$67.39$ & \cellcolor{MyColor0!50!white}$67.17$ & \cellcolor{MyColor2!50!white}$67.27$ & \cellcolor{MyColor0!50!white}$67.19$ & \cellcolor{MyColor0!50!white}$67.16$ & \cellcolor{MyColor0!50!white}$67.15$ & \cellcolor{MyColor0!50!white}$67.15$ & \cellcolor{MyColor0!50!white}$67.14$ & \cellcolor{MyColor0!50!white}$67.14$ & \cellcolor{MyColor0!50!white}$67.14$ & \cellcolor{MyColor0!50!white}$67.15$ & \cellcolor{MyColor0!50!white}$67.15$ & \cellcolor{MyColor0!50!white}$67.16$ & \cellcolor{MyColor0!50!white}$67.17$ \\\hline
$k = 7$ & \cellcolor{MyColor76!50!white}$68.49$ & \cellcolor{MyColor26!50!white}$67.61$ & \cellcolor{MyColor0!50!white}$67.21$ & \cellcolor{MyColor0!50!white}$67.18$ & \cellcolor{MyColor0!50!white}$67.07$ & \cellcolor{MyColor0!50!white}$67.01$ & \cellcolor{MyColor0!50!white}$66.98$ & \cellcolor{MyColor0!50!white}$66.97$ & \cellcolor{MyColor0!50!white}$66.97$ & \cellcolor{MyColor0!50!white}$66.97$ & \cellcolor{MyColor0!50!white}$66.97$ & \cellcolor{MyColor0!50!white}$66.98$ & \cellcolor{MyColor0!50!white}$66.98$ & \cellcolor{MyColor0!50!white}$66.99$ & \cellcolor{MyColor0!50!white}$67.00$ & \cellcolor{MyColor0!50!white}$67.01$ \\\hline
$k = 8$ & \cellcolor{MyColor59!50!white}$68.16$ & \cellcolor{MyColor14!50!white}$67.43$ & \cellcolor{MyColor0!50!white}$67.09$ & \cellcolor{MyColor0!50!white}$67.05$ & \cellcolor{MyColor0!50!white}$66.98$ & \cellcolor{MyColor0!50!white}$\mathbf{66.93}$ & \cellcolor{MyColor0!50!white}$\mathbf{66.92}$ & \cellcolor{MyColor0!50!white}$\mathbf{66.93}$ & \cellcolor{MyColor0!50!white}$66.94$ & \cellcolor{MyColor0!50!white}$66.94$ & \cellcolor{MyColor0!50!white}$66.96$ & \cellcolor{MyColor0!50!white}$66.97$ & \cellcolor{MyColor0!50!white}$66.99$ & \cellcolor{MyColor0!50!white}$67.00$ & \cellcolor{MyColor0!50!white}$67.02$ & \cellcolor{MyColor0!50!white}$67.04$ \\\hline
$k = 9$ & \cellcolor{MyColor39!50!white}$67.82$ & \cellcolor{MyColor2!50!white}$67.27$ & \cellcolor{MyColor0!50!white}$66.98$ & \cellcolor{MyColor0!50!white}$66.99$ & \cellcolor{MyColor0!50!white}$66.96$ & \cellcolor{MyColor0!50!white}$66.94$ & \cellcolor{MyColor0!50!white}$66.96$ & \cellcolor{MyColor0!50!white}$66.99$ & \cellcolor{MyColor0!50!white}$67.01$ & \cellcolor{MyColor0!50!white}$67.03$ & \cellcolor{MyColor0!50!white}$67.06$ & \cellcolor{MyColor0!50!white}$67.09$ & \cellcolor{MyColor0!50!white}$67.12$ & \cellcolor{MyColor0!50!white}$67.15$ & \cellcolor{MyColor0!50!white}$67.18$ & \cellcolor{MyColor0!50!white}$67.21$ \\\hline
$k = 10$ & \cellcolor{MyColor33!50!white}$67.72$ & \cellcolor{MyColor10!50!white}$67.38$ & \cellcolor{MyColor0!50!white}$67.13$ & \cellcolor{MyColor0!50!white}$67.20$ & \cellcolor{MyColor0!50!white}$67.21$ & \cellcolor{MyColor0!50!white}$67.22$ & \cellcolor{MyColor1!50!white}$67.25$ & \cellcolor{MyColor4!50!white}$67.30$ & \cellcolor{MyColor7!50!white}$67.34$ & \cellcolor{MyColor10!50!white}$67.38$ & \cellcolor{MyColor13!50!white}$67.42$ & \cellcolor{MyColor16!50!white}$67.47$ & \cellcolor{MyColor19!50!white}$67.52$ & \cellcolor{MyColor22!50!white}$67.56$ & \cellcolor{MyColor25!50!white}$67.60$ & \cellcolor{MyColor28!50!white}$67.65$ \\\hline
$k = 11$ & \cellcolor{MyColor52!50!white}$68.04$ & \cellcolor{MyColor40!50!white}$67.83$ & \cellcolor{MyColor31!50!white}$67.69$ & \cellcolor{MyColor30!50!white}$67.67$ & \cellcolor{MyColor32!50!white}$67.71$ & \cellcolor{MyColor33!50!white}$67.73$ & \cellcolor{MyColor36!50!white}$67.77$ & \cellcolor{MyColor40!50!white}$67.84$ & \cellcolor{MyColor43!50!white}$67.89$ & \cellcolor{MyColor47!50!white}$67.95$ & \cellcolor{MyColor51!50!white}$68.02$ & \cellcolor{MyColor54!50!white}$68.08$ & \cellcolor{MyColor58!50!white}$68.15$ & \cellcolor{MyColor61!50!white}$68.21$ & \cellcolor{MyColor65!50!white}$68.27$ & \cellcolor{MyColor68!50!white}$68.33$ \\\hline
$k = 12$ & \cellcolor{MyColor79!50!white}$68.54$ & \cellcolor{MyColor64!50!white}$68.26$ & \cellcolor{MyColor62!50!white}$68.22$ & \cellcolor{MyColor63!50!white}$68.25$ & \cellcolor{MyColor68!50!white}$68.33$ & \cellcolor{MyColor72!50!white}$68.40$ & \cellcolor{MyColor76!50!white}$68.49$ & \cellcolor{MyColor81!50!white}$68.59$ & \cellcolor{MyColor86!50!white}$68.69$ & \cellcolor{MyColor91!50!white}$68.79$ & \cellcolor{MyColor95!50!white}$68.88$ & \cellcolor{MyColor99!50!white}$68.98$ & \cellcolor{MyColor103!50!white}$69.07$ & \cellcolor{MyColor107!50!white}$69.15$ & \cellcolor{MyColor110!50!white}$69.24$ & \cellcolor{MyColor114!50!white}$69.32$ \\\hline
$k = 13$ & \cellcolor{MyColor87!50!white}$68.72$ & \cellcolor{MyColor90!50!white}$68.78$ & \cellcolor{MyColor97!50!white}$68.94$ & \cellcolor{MyColor103!50!white}$69.07$ & \cellcolor{MyColor110!50!white}$69.23$ & \cellcolor{MyColor115!50!white}$69.36$ & \cellcolor{MyColor121!50!white}$69.50$ & \cellcolor{MyColor126!50!white}$69.64$ & \cellcolor{MyColor131!50!white}$69.77$ & \cellcolor{MyColor135!50!white}$69.90$ & \cellcolor{MyColor139!50!white}$70.02$ & \cellcolor{MyColor143!50!white}$70.13$ & \cellcolor{MyColor146!50!white}$70.24$ & \cellcolor{MyColor149!50!white}$70.34$ & \cellcolor{MyColor152!50!white}$70.44$ & \cellcolor{MyColor155!50!white}$70.53$ \\\hline
$k = 14$ & \cellcolor{MyColor109!50!white}$69.21$ & \cellcolor{MyColor125!50!white}$69.62$ & \cellcolor{MyColor137!50!white}$69.95$ & \cellcolor{MyColor144!50!white}$70.16$ & \cellcolor{MyColor150!50!white}$70.36$ & \cellcolor{MyColor155!50!white}$70.54$ & \cellcolor{MyColor159!50!white}$70.71$ & \cellcolor{MyColor164!50!white}$70.86$ & \cellcolor{MyColor167!50!white}$71.00$ & \cellcolor{MyColor170!50!white}$71.14$ & \cellcolor{MyColor173!50!white}$71.26$ & \cellcolor{MyColor176!50!white}$71.38$ & \cellcolor{MyColor178!50!white}$71.49$ & \cellcolor{MyColor180!50!white}$71.59$ & \cellcolor{MyColor182!50!white}$71.68$ & \cellcolor{MyColor184!50!white}$71.77$ \\\hline
$k = 15$ & \cellcolor{MyColor141!50!white}$70.07$ & \cellcolor{MyColor163!50!white}$70.82$ & \cellcolor{MyColor170!50!white}$71.12$ & \cellcolor{MyColor176!50!white}$71.36$ & \cellcolor{MyColor181!50!white}$71.60$ & \cellcolor{MyColor184!50!white}$71.76$ & \cellcolor{MyColor187!50!white}$71.92$ & \cellcolor{MyColor190!50!white}$72.05$ & \cellcolor{MyColor192!50!white}$72.18$ & \cellcolor{MyColor255!50!white}n/a & \cellcolor{MyColor255!50!white}n/a & \cellcolor{MyColor255!50!white}n/a & \cellcolor{MyColor255!50!white}n/a & \cellcolor{MyColor255!50!white}n/a & \cellcolor{MyColor255!50!white}n/a & \cellcolor{MyColor255!50!white}n/a \\\hline
$k = 16$ & \cellcolor{MyColor172!50!white}$71.19$ & \cellcolor{MyColor190!50!white}$72.10$ & \cellcolor{MyColor195!50!white}$72.35$ & \cellcolor{MyColor198!50!white}$72.54$ & \cellcolor{MyColor201!50!white}$72.70$ & \cellcolor{MyColor203!50!white}$72.86$ & \cellcolor{MyColor255!50!white}n/a & \cellcolor{MyColor255!50!white}n/a & \cellcolor{MyColor255!50!white}n/a & \cellcolor{MyColor255!50!white}n/a & \cellcolor{MyColor255!50!white}n/a & \cellcolor{MyColor255!50!white}n/a & \cellcolor{MyColor255!50!white}n/a & \cellcolor{MyColor255!50!white}n/a & \cellcolor{MyColor255!50!white}n/a & \cellcolor{MyColor255!50!white}n/a \\\hline
$k = 17$ & \cellcolor{MyColor195!50!white}$72.37$ & \cellcolor{MyColor207!50!white}$73.16$ & \cellcolor{MyColor255!50!white}n/a & \cellcolor{MyColor255!50!white}n/a & \cellcolor{MyColor255!50!white}n/a & \cellcolor{MyColor255!50!white}n/a & \cellcolor{MyColor255!50!white}n/a & \cellcolor{MyColor255!50!white}n/a & \cellcolor{MyColor255!50!white}n/a & \cellcolor{MyColor255!50!white}n/a & \cellcolor{MyColor255!50!white}n/a & \cellcolor{MyColor255!50!white}n/a & \cellcolor{MyColor255!50!white}n/a & \cellcolor{MyColor255!50!white}n/a & \cellcolor{MyColor255!50!white}n/a & \cellcolor{MyColor255!50!white}n/a \\\hline
\end{tabularx}
\end{small}
\caption{Size of compressed lists for full TREC dataset with adaptive contexts. All numbers are given $\pm 0.01$~MB. Cells indicating ``n/a'' are cells where the memory size of the occurrences exceeded 15~GB of RAM.}
\end{table}
\end{landscape}

\newpage

\small
\defbibfilter{all}{
	type=book or
	type=thesis or
	type=incollection or
	type=article or
	type=inproceedings or
	type=report
}
\defbibfilter{tools}{
	type=manual
}
\defbibfilter{inverted_index}{
	( type=book or
	type=thesis or
	type=incollection or
	type=article or
	type=inproceedings or
	type=report )
	and
	( keyword=compression or keyword=reordering )
}
\defbibfilter{encodings}{
	( type=book or
	type=thesis or
	type=incollection or
	type=article or
	type=inproceedings or
	type=report )
	and
	( keyword=encoding or keyword=entropy )
}
\AtNextBibliography{\footnotesize}{\tiny }
\printbibliography[heading=subbibliography,title={Books and articles --- Inverted index compression techniques},filter=inverted_index,notkeyword={dataset},notkeyword={encoding},notkeyword={entropy},notkeyword={software}]
\AtNextBibliography{\footnotesize}
\printbibliography[heading=subbibliography,title={Books and articles --- Encoding},filter=encodings,notkeyword={compression},notkeyword={dataset},notkeyword={reordering},notkeyword={software}]
\AtNextBibliography{\footnotesize}
\printbibliography[heading=subbibliography,title={Books and articles --- Other},filter=all,notkeyword={compression},notkeyword={dataset},notkeyword={encoding},notkeyword={entropy},notkeyword={reordering},notkeyword={software}]
\AtNextBibliography{\footnotesize}
\printbibliography[heading=subbibliography,title={Datasets},keyword={dataset},notkeyword={compression},notkeyword={encoding},notkeyword={entropy},notkeyword={reordering},notkeyword={software}]
\AtNextBibliography{\footnotesize}
\printbibliography[heading=subbibliography,title={Softwares},keyword={software},notkeyword={dataset}]

\end{document}